\documentclass[
	twocolumn,
    tighten,
	]{aastex63}

\usepackage{bm}
\usepackage{amsmath}
\usepackage{amssymb}
\usepackage{color}

\usepackage[T1]{fontenc}
\usepackage{newpxtext, mathptmx}

\renewcommand{\eqref}[1]{Eq.~(\ref{#1})}
\newcommand{\figref}[1]{Fig.~\ref{#1}}

\newcommand{\Alfven}{Alfv\'en }

\newcommand{\MAHTF}{M_{\rm A}^{\rm HTF}}
\newcommand{\MA}{M_{\rm A}}
\newcommand{\VA}{V_{\rm A}}
\newcommand{\thetabn}{\theta_{Bn}}

\newcommand{\wci}{\Omega_{\rm ci}}
\newcommand{\wce}{\Omega_{\rm ce}}
\newcommand{\wpi}{\omega_{\rm pi}}

\newcommand{\betai}{\beta_{\rm i}}
\newcommand{\betae}{\beta_{\rm e}}
\newcommand{\rg}{r_{\rm g}}
\newcommand{\mfp}{\lambda_{\rm mfp}}
\newcommand{\ldiff}{l_{\rm diff}}
\newcommand{\pae}{p_{\rm A, e}}

\newcommand{\UP}{\rm u}
\newcommand{\DOWN}{\rm d}

\submitjournal{ApJ}

\shorttitle{Electron Injection at Oblique Shock}
\shortauthors{Amano \& Hoshino}

\begin{document}

\title{Theory of Electron Injection at Oblique Shock of Finite Thickness}

\correspondingauthor{Takanobu Amano}
\email{amano@eps.s.u-tokyo.ac.jp}
\author[0000-0002-2140-6961]{Takanobu Amano}
\affiliation{Department of Earth and Planetary Science, University of Tokyo \\
7-3-1 Hongo, Bunkyo-ku, Tokyo, 113-0033, Japan}

\author[0000-0002-1818-9927]{Masahiro Hoshino}
\affiliation{Department of Earth and Planetary Science, University of Tokyo \\
7-3-1 Hongo, Bunkyo-ku, Tokyo, 113-0033, Japan}



\begin{abstract}
A theory of electron injection into diffusive shock acceleration (DSA) for the generation of cosmic-ray electrons at collisionless shocks is presented. We consider a recently proposed particle acceleration mechanism called stochastic shock drift acceleration (SSDA). We find that SSDA may be understood as a diffusive particle acceleration mechanism at an oblique shock of finite thickness. More specifically, it is described by a solution to the diffusion-convection equation for particles with the characteristic diffusion length comparable to the shock thickness. On the other hand, the same equation yields the standard DSA if the diffusion length is much longer than the thickness. Although SSDA predicts, in general, a spectral index steeper than DSA, it is much more efficient for low-energy electron acceleration and is favorable for injection. The injection threshold energy corresponds to the transition energy between the two different regimes. It is of the order of $0.1\text{--}1$ MeV in typical interstellar and interplanetary conditions if the dissipation scale of turbulence around the shock is determined by the ion inertial length. The electron injection is more efficient at high $\MA / \cos \thetabn$ where $\MA$ and $\thetabn$ are the \Alfven Mach number and the shock obliquity. The theory suggests that efficient acceleration of electrons to ultra-relativistic energies will be more easily realized at high-Mach-number young supernova remnant shocks, but not at weak or moderate shocks in the heliosphere unless the upstream magnetic field is nearly perpendicular to the shock normal.
\end{abstract}



\section{Introduction}
\label{sec:intro}
The acceleration of charged particles to energies well beyond the thermal energy is ubiquitous in collisionless astrophysical and heliospheric plasmas. The rarity of binary Coulomb collisions in these environments allows non-thermal particles to survive without losing energies for a long time. It is, however, not well understood how such non-thermal particles emerge out of the thermal population.

One of the most well-known mechanisms of particle acceleration is associated with shock waves. It is widely accepted that shock waves in supernova remnants (SNRs) are the primary sources of galactic cosmic rays (CRs). Radio and X-ray synchrotron emissions from ultra-relativistic electrons in young SNRs support the standard paradigm \citep[e.g.,][]{Reynolds2008}. While the acceleration of electrons to ultra-relativistic energies appears to be common in strong astrophysical shocks, relativistic electrons are rarely observed around shocks in the heliosphere \citep{Lario2003,Dresing2016}. In-situ observations of Earth's bow shock, the best-studied example of collisionless shocks, indicate that the highest electron energy is limited to a few hundreds of keV even in ``efficient'' events \citep{Anderson1981,Gosling1989,Wilson2016}. Although it may partly be due to limited energy ranges and sensitivities of the measurements, it appears that there is an intrinsic limit of the electron acceleration efficiency. In this paper, we will focus on the reason why such a difference arises depending on the shock parameters. We note that, although the proton acceleration efficiency at SNRs should be the major interest in resolving the CR origins, understanding the electron acceleration is crucial as they provide valuable information of remote particle acceleration sites with their higher radiative efficiency.

The diffusive shock acceleration (DSA) mechanism has been considered as the standard model of particle acceleration at shocks \citep{Bell1978a,Blandford1978a,Drury1983,Blandford1987}, which has achieved considerable success over the decades. It assumes that the particle transport around the shock is described by diffusion, and the particle may freely traverse the shock. The diffusion results from elastic scattering of particles by magnetic fluctuations convected by the background plasma flow. The scattering thus brings an energy gain and loss in the upstream and downstream, respectively. Since the flow in the upstream is always faster than the downstream, there is a net energy gain every time the particle bounces back and forth across the shock front. One of the advantages of the theory is that it predicts a nearly universal power-law energy spectrum at strong shocks, which is independent of the details of scattering. The predicted spectral index is roughly consistent with various observations. A well-known drawback, known as the injection problem, is that DSA requires a sufficiently energetic seed population that is assumed to be provided by some other mechanisms. In other words, since it does not provide a prediction on the low-energy part of the spectrum, it is impossible to know the absolute amount of CRs. In particular, the injection of electrons is known to be notoriously more difficult than that of protons and heavier nuclei for reasons that will be discussed later.

Extensive effort has been devoted to trying to resolve the long-standing issue of electron injection. Since the acceleration of low-energy, supra-thermal electrons is intimately linked to the shock dissipation itself, the excitation of microscopic plasma waves and their roles on particle energization have been topics of great interest. The fully kinetic Particle-in-Cell (PIC) simulation has been the most powerful tool for the purpose. Starting from early simulations in one dimension (1D) \citep[e.g.,][]{Shimada2000,Hoshino2002,Scholer2003a}, recent simulations have been performed routinely in two dimensions (2D) \citep[e.g.,][]{Amano2009a,Umeda2012b,Tran2020,Bohdan2021} and sometimes even in three dimensions (3D) \citep{Matsumoto2017}. Similarly, in-situ measurements by spacecraft, in particular at Earth's bow shock, have been used to study the electron dynamics at collisionless shocks \citep{Feldman1982a,Feldman1983a,Thomsen1987a,Scudder1986a,Schwartz1988b}. Various kinds of waves (both electrostatic and electromagnetic) have been found, and their roles on particle heating have been discussed in detail \citep[e.g.,][]{Wilson2014}. Nevertheless, the time resolution of particle measurements had not been able to fully resolve the shock structure until the launch of Magnetospheric Multi-Scale (MMS) spacecraft. By using MMS observations, \citet{Oka2017} found clear evidence of the interaction between low-energy electrons and high-frequency whistler-mode waves via cyclotron resonance within the shock transition layer. On the simulation side, 3D PIC simulations for a high-Mach-number quasi-perpendicular shock found that the electron energization within the shock transition layer was associated with strong scatterings in pitch angle \citep{Matsumoto2017}.

Motivated by these findings, \citet{Katou2019} proposed a new particle acceleration mechanism which we call stochastic shock drift acceleration (SSDA). In short, SSDA is essentially the conventional shock drift acceleration (SDA) mechanism \citep[e.g.,][]{Wu1984b,Leroy1984a} under the presence of sufficiently strong pitch-angle scattering. \citet{Amano2020} recently demonstrated a good agreement between the SSDA theory and in-situ MMS spacecraft observations at Earth's bow shock. While Earth's bow shock is not an efficient site of electron acceleration to relativistic energies, the theory indicates that SSDA will be able to inject electrons to DSA at young SNRs for further energization to ultra-relativistic energies. To the authors' knowledge, SSDA is a unique electron injection mechanism proposed thus far in that the model has been tested using in-situ spacecraft observations. Furthermore, the apparent discrepancy between astrophysical and heliospheric shocks may reasonably be explained with its theoretical scaling law. We thus believe that, at present, SSDA is the most promising candidate for electron injection, which deserves further scrutiny.

In this paper, we extend the existing theory of SSDA for more quantitative modeling of electron acceleration at collisionless shocks. We use a diffusion-convection equation in the de Hoffmann-Teller frame (HTF), which is consistently derived from a general fully-relativistic transport equation under the assumption of weak anisotropy (see, Appendix \ref{sec:transport-equation}). With the diffusion-convection equation, we consider particle acceleration at an oblique shock of finite thickness. We find that the solution naturally reduces to the standard DSA when the particle diffusion length is much longer than the thickness of the shock. On the other hand, SSDA corresponds to the solution in which the diffusion length is comparable to the thickness. In other words, both DSA and SSDA may be obtained as solutions of the same equation in different limiting cases. This suggests that both the pre-acceleration by SSDA (before the injection) and further acceleration by DSA (after the injection) may be described by the present model in a consistent fashion. The injection in this scenario is essentially the transition between the two different regimes. Beyond the transition energy, electrons may be accelerated via DSA even if their Larmor radii are smaller than the shock thickness. As we shall see, we estimate the injection threshold energy assuming that the turbulence around the shock provides the particle scattering. In other words, the fluctuation power at wavelength comparable to the Larmor radii of the particles being accelerated needs to be sufficiently large. If the turbulence dissipation scale is determined by the ion inertial length, the threshold energy may be on the order of $0.1\text{--}1$ MeV in typical interstellar and interplanetary conditions. An efficient injection will thus be realized only if SSDA is able to accelerate particles beyond the threshold energy. Although the condition depends on unknown parameters (see, \eqref{eq:ma-cond-injection}), we conclude that shocks with higher \Alfven Mach numbers in HTF ($\MAHTF$) will be more efficient sites of electron injection. We think that the theory developed in this paper provides the basis for understanding the electron injection problem.


The present paper will be organized as follows. Based on the diffusion-convection equation derived in Appendix \ref{sec:transport-equation}, analytic theory for electron injection will be presented in Section \ref{sec:electron-injection}, where we show that both DSA and SSDA are obtained from the same equation. The transition between the two and the injection threshold energy will also be described. In Section \ref{sec:numerical-solutions}, we discuss fully numerical solutions of the diffusion-convection equation. We show that, indeed, the electron acceleration regime changes from SSDA at low energy to DSA at high energy across the transition energy that is determined by the ratio between the diffusion length and the shock thickness. Section \ref{sec:discussion} is devoted to discussion of the present results and, in particular, perspectives for future observational (both in-situ and remote-sensing), experimental, and numerical investigations. Finally, Section \ref{sec:summary} summarizes the paper.

\section{Theory of Electron Injection}
\label{sec:electron-injection}

\subsection{Reference Frames}
A natural frame of reference to analyze a magnetized oblique shock wave is the so-called normal incident frame (NIF), in which the shock is at rest and the upstream flow is parallel to the shock normal. We define the shock speed as the upstream flow speed $U_{\rm s}$ in the NIF. The magnetic obliquity $\thetabn$ is defined as the angle between the upstream magnetic field and the shock normal direction. In the following, we will always work in HTF, in which the plasma flow and the magnetic field are parallel with each other. Since the motional electric field vanishes in this special frame, it is one of the most convenient frames to analyze the transport of charged particles around a collisionless oblique shock. Such a frame may be found by frame transformation in the direction transverse to the shock with a speed of $U_{\rm s} \tan \thetabn$. The transformation speed must obviously be less than the speed of light $U_{\rm s} \tan \thetabn < c$ (where $c$ is the speed of light). Such a shock is called subluminal. On the other hand, if $U_{\rm s} \tan \thetabn \geq c$, we cannot find HTF, and the shock is called superluminal. In this paper, we will consider only subluminal shocks because most of the non-relativistic oblique shocks of practical interest are subluminal. For instance, $\thetabn < 89 \degr$ is sufficient for a young SNR shock of a shock speed of $U_{\rm s} = 3000 \, {\rm km/s}$ to be subluminal. Appendix \ref{sec:transport-equation} shows that the transport equation conventionally written in an arbitrary frame can be cast into a simpler form in HTF. As in the case of single-particle dynamics, the explicit use of HTF makes it easy to understand the physics involved.

In the following, we will denote the flow velocity in HTF by $\bm{V}$. It is, by definition, always parallel to the local magnetic field, i.e., $\bm{V} = V \bm{b}$ where $\bm{b}$ is the magnetic field unit vector. The flow velocity in NIF is denoted by $\bm{U}$.

\subsection{Basic Equation}
We investigate the electron injection at a plane oblique shock of finite thickness of order $L_{\rm s}$ by considering the steady-state solution (i.e., $\partial/\partial t = 0$) of the diffusion-convection equation
\begin{align}
	\frac{\partial f}{\partial t} +
	V \cos \theta \, \frac{\partial f}{\partial x} +
	\frac{1}{3}
	\left(
		\frac{\partial \ln B}{\partial x} - \frac{\partial \ln V}{\partial x}
	\right)
	V \cos \theta \frac{\partial f}{\partial \ln p} \nonumber \\
    =
	\frac{\partial}{\partial x}
	\left(
		\kappa \cos^2 \theta \, \frac{\partial f}{\partial x}
	\right),
	\label{eq:diffusion-transport}
\end{align}
where $f = f(x, p)$ is the isotropic part of the distribution function as a function of position $x$ and plasma-frame momentum $p$. The meanings of other quantities are as follows: $B = B(x)$ is the magnetic field strength, $\theta = \theta(x)$ is the angle between the magnetic field and $x$ axis, $V = V(x)$ is the field-aligned flow speed, $\kappa = \kappa(x, p)$ is the diffusion coefficient along the local magnetic field. Note that \eqref{eq:diffusion-transport} is equivalent to \eqref{eq:diffusion-transport-appendix} except that we have dropped the subscript for $f_0$ as we will consider only the isotropic part throughout this paper (see, Appendix \ref{sec:transport-equation} for detail). We assume that the shock normal is parallel to the $x$ axis and the flow velocity and the magnetic field strength change smoothly from the upstream values $V_1, B_1, \theta_1 (= \thetabn)$ at $x \rightarrow -\infty$ to the downstream values $V_2, B_2, \theta_2$ at $x \rightarrow +\infty$. Note that the upstream flow speed in HTF is given by $V_1 = U_{\rm s} / \cos \thetabn$.

We find it convenient to use the following form to estimate the spectral index:
\begin{align}
	\int_{X_1}^{X_2}
	\frac{\partial V_x}{\partial x}
	\left[ f + \frac{1}{3} \frac{\partial f}{\partial \ln p} \right]
	dx =
	\left[
		V_x f - \kappa_{xx} \frac{\partial f}{\partial x}
	\right]_{X_1}^{X_2},
	\label{eq:index-formula}
\end{align}
which is obtained by integrating \eqref{eq:diffusion-transport} with respect to $x$ on an arbitrary interval $[X_1, X_2]$. Note that we have introduced the notations: $V_x = V \cos \theta$ and $\kappa_{xx} = \kappa \cos^2 \theta$.

It is easy to find from \eqref{eq:diffusion-transport} that the characteristic scale length is given by $\ldiff = \kappa_{xx}/V_x = \kappa \cos^2 \theta/U_{\rm s}$, which we call the diffusion length. More specifically, the spatial profile is given by $f(x) \propto \exp \left( x / \ldiff \right)$ for a system that is approximately homogeneous. We thus find that the particle acceleration at a shock of finite thickness can be classified by the ratio between the diffusion length and the shock thickness $\ldiff / L_{\rm s}$.

\subsection{Diffusive Shock Acceleration}
\label{sec:dsa}

We first look at the conventional DSA as an obvious example. If the diffusion length for a paricular particle energy is much larger than the thickness of the shock $\ldiff/L_{\rm s} \gg 1$, the shock as seen by the particle is essentially a discontinuity in the plasma flow:
\begin{align}
	\frac{\partial V_x}{\partial x} \approx
	\left( V_2 \cos \theta_2 - V_1 \cos \theta_1 \right) \, \delta (x).
\end{align}

We adopt the boundary condition such that $f = 0$ at $x = X_1$ and $\partial f/\partial x = 0$ at $x = X_2$. Carrying out the integral in \eqref{eq:index-formula} in the limit of infinite integration interval $X_1 \rightarrow -\infty$ and $X_2 \rightarrow +\infty$, we have
\begin{align}
	\left( V_1 \cos \theta_1 - V_2 \cos \theta_2 \right)
	\left[ f_2 + \frac{1}{3} \frac{\partial f_2}{\partial \ln p} \right]
	=
	V_2 \cos \theta_2 f_2,
\end{align}
where $f_2 = f(x=X_2, p)$. It becomes immediately clear that the spectrum becomes a power-law $f_2(p) \propto p^{-q}$ with its index given by
\begin{align}
	q = \frac{3 V_1 \cos \theta_1}{V_2 \cos \theta_2 - V_1 \cos \theta_1}
	= \frac{3 r}{r - 1},
\end{align}
where $r = V_1 \cos \theta_1 / V_2 \cos \theta_2 = V_{x,1}/V_{x,2}$ is the compression ratio. It is well known that the spectral index of particles accelerated at an oblique shock is determined solely by the compression ratio, and independent on the magnetic obliquity $\thetabn$ even though the energy gain clearly includes both the first-order Fermi acceleration and SDA effects \citep[e.g.,][]{Drury1983}.

\subsection{Stochastic Shock Drift Acceleration}
\label{sec:ssda}

Let us now consider the general case in which the effect of a finite shock thickness needs to be taken into account. As in the case of the discontinuous shock solution, we may anticipate the particle density increases within the shock from the far upstream value and then becomes constant in the downstream. We thus use the same boundary condition as DSA. By assuming the solution of the form $f(x, p) = g(x) p^{-q}$, we estimate the spectral index by
\begin{align}
	q = 3
	\left[
		1 -
		\left(
			\int_{X_1}^{X_2} \frac{g(x)}{g_2}
			\frac{\partial}{\partial x}
			\left( \frac{V_x}{V_{x,2}} \right)
			dx
		\right)^{-1}
	\right].
	\label{eq:general-index-formula}
\end{align}
where $g_2 = g(x = X_2)$. Note that the integration interval $[X_1, X_2]$ is arbitrary as long as it contains the entire particle acceleration region. It is easy to see that the canonical DSA spectral index may be recovered for a discontinuous flow profile. For more general cases, the evaluation of the integral requires a specific form of $g(x)$ that will only be known by solving \eqref{eq:diffusion-transport}. Nevertheless, this formula is useful for comparisons with in-situ observations and kinetic simulations because in both cases, the profiles of $g(x)$ and $V_x(x)$ are already known.

It is instructive to proceed further to obtain an analytic estimate for the spectral index without solving \eqref{eq:diffusion-transport}. By looking at fully numerical solutions (see, Section \ref{sec:numerical-solutions}), we have found that the exponential profile $g(x) \propto \exp(x/\ldiff)$ is approximately correct even within the shock transition layer $\partial V_x/\partial x \neq 0$. Since the condition $\ldiff/L_{\rm s} = \kappa \cos \thetabn^2 /U_{\rm s} \lesssim 1$ favors quasi-perpendicular shocks, we here only consider particle acceleration at a finite-thickness quasi-perpendicular shock, which we call SSDA. In this case, we may ignore the particle energy changes due to the flow compression ($V_1 \approx V_2$) and the kink in the magnetic field line ($\theta_1 \approx \theta_2$) at quasi-perpendicular shocks. We then have
\begin{align}
	\frac{\partial V_x}{\partial x} \approx -
	V_1 \cos \theta_1 \frac{\partial \ln B}{\partial x}.
\end{align}
If we further adopt an average magnetic field gradient $\partial \ln B/\partial x \approx \left< \partial \ln B/\partial x \right> \approx {\rm const.}$ to evaluate the integral, we finally obtain the following estimate for the spectral index:
\begin{align}
	q \approx 3
	\left[
		1 +
		\left(
			\ldiff \left< \frac{\partial \ln B}{\partial x} \right>
		\right)^{-1}
	\right].
	\label{eq:ssda-approx-index}
\end{align}
It is determined by the ratio between the diffusion length to the magnetic field transition scale length $\left( \partial \ln B / \partial x \right)^{-1} \sim L_{\rm s}$. Since, in general, the diffusion coefficient is energy-dependent, the spectrum of the accelerated particles will not be described by a single power law. A power-law spectrum will be obtained only when the diffusion length $\ldiff$ (and thus the diffusion coefficient $\kappa$) is independent of energy.

It is easy to see that the spectral index for SSDA is steeper than the DSA prediction. Since, by definition, the diffusion length should be smaller than the shock thickness $\ldiff \lesssim L_{\rm s}$, it should have a lower limit $q \gtrsim 6$. Even the hardest spectrum predicted for SSDA is softer than $q = 4$ predicted by DSA. The spectrum becomes even softer as the diffusion length decreases.

The steep spectrum for particles with small diffusion lengths is a natural consequence of efficient particle confinement. In other words, the convective escape time of particles $\tau_{\rm esc} \approx r ( \ldiff/U_{\rm s} )$ from the acceleration region becomes shorter as the confinement efficiency increases, while the acceleration time scale $\tau_{\rm acc} \approx 3 L_{\rm s}/U_{\rm s}$ remains unchanged. This is in clear contrast to DSA in which $\tau_{\rm acc} \approx 3 r/(r-1) ( \ldiff/U_{\rm s} )$ has the same dependence on $\ldiff$. This is indeed the reason why DSA gives the nearly universal power-law that is independent on the diffusion coefficient: $q = \tau_{\rm acc}/\tau_{\rm esc} + 3 = 3 r/(r - 1)$ \citep[see, e.g.,][]{Drury1991}.

It is important to point out that SSDA is a fast process instead. Since the particle acceleration via SSDA operates locally within the shock transition layer, there is no need for particles bouncing back and forth across the shock, thereby saving time. Note that high-frequency waves, which play the role of scattering of low-energy electrons, are most intense in the close vicinity of the shock \citep[e.g.,][]{Hull2012,Wilson2014,Oka2017,Amano2020}. Therefore, it does not require high-frequency wave activity far from the shock, where the waves will be subject to strong collisionless damping. Considering that the diffusion length may, in general, be considered as a monotonically increasing function of energy, one may expect that SSDA and DSA tend to be dominant mechanisms for particle acceleration at low-energy and high-energy, respectively.

We should mention that the spectral index \eqref{eq:ssda-approx-index} is similar to, but different from the estimate given by \citet{Katou2019} (in which the spectral index is defined for $N(p) = 4 \pi p^2 f(p) \propto p^{-q + 2}$ for non-relativistic particles). More specifically, $\ldiff$ in \eqref{eq:ssda-approx-index} is replaced by the typical shock thickness $L_{\rm s}$ in \citet{Katou2019}. This difference arises because they estimated the typical escape time by $\tau_{\rm esc} \approx L_{\rm s}/U_{\rm s}$, i.e., the convection time scale over the shock thickness $L_{\rm s}$, rather than the diffusion length $\ldiff$. The erroneous expression should be considered appropriate only when the diffusion length is energy-independent and comparable to the shock thickness. Interestingly enough, in-situ measurements at Earth's bow shock suggested that both of these conditions were reasonably satisfied in the energy range where the spectrum is described by a power law \citep{Amano2020}, which is fully consistent with the theory.

\subsection{Transition from SSDA to DSA}
\label{sec:ssda-to-dsa}

As we have seen in previous subsections, the acceleration of particles for a given diffusion coefficient may be understood in terms of DSA or SSDA. So far, we have developed the general discussion without specifying the shock thickness, particle species, and particle energy. From now on, we consider the specific application to the electron injection problem.

It is well known that the transition layer thickness of super-critical collisionless shocks is of the order of the ion gyroradius $U_{\rm s}/\wci$, where $\wci$ is the ion cyclotron frequency in the upstream \citep[e.g.,][]{Livesey1984}. Note that the thickness always refers to the length scale of the shock transition layer including the foot, ramp, and overshoot in this paper. We thus define the thickness by $L_{\rm s} = \eta U_{\rm s}/\wci$ with $\eta$ being a numerical factor of order unity. The ratio between the diffusion length to the shock thickness is thus given by:
\begin{align}
	\frac{\ldiff}{L_{\rm s}} =
	\frac{1}{6 \eta} \left( \frac{D_{\mu\mu}}{\wci} \right)^{-1}
	\left( \frac{v}{U_{\rm s}/\cos \thetabn} \right)^{2}.
\end{align}
To evaluate the quantity, we may consider that the minimum velocity of the particles being accelerated corresponds to that expected by adiabatic SDA $v \gtrsim U_{\rm s}/\cos \thetabn$. Since the scattering rate will be given at most by the Bohm limit, ions with $D_{\mu\mu}/\wci \lesssim 1$ should not be able to satisfy the condition unless the shock is excessively thicker than normal ($\eta \gg 1$). On the other hand, the scattering of electrons can be much faster $D_{\mu\mu}/\wci \lesssim m_i/m_e$, where $m_i/m_e$ is the ion-to-electron mass ratio. Therefore, low-energy electrons may be subject to particle acceleration via SSDA within the shock transition layer.

Consider a low-energy electron that is being accelerated by SSDA. Because of the upper limit on the scattering rate, there should be the maximum energy $E_{\rm max, SSDA}$ that can be achievable through SSDA. This may be determined by the condition $\ldiff/L_{\rm s} \simeq 1$, yielding
\begin{align}
	\frac{E_{\rm max, SSDA}}{E_{0}}
	\simeq 6 \eta \left( \frac{D_{\mu\mu}}{\wci} \right),
	\label{eq:emax-ssda}
\end{align}
where $E_{0} = m_e (U_{\rm s}/\cos \thetabn)^2 / 2$. Note that this expression is appropriate only for non-relativistic electrons. It is clear that electrons beyond this energy cannot be trapped inside the shock transition layer but may potentially be accelerated even further via DSA in a much larger spatial extent. In other words, one may expect that a steep spectrum will develop at low energy $E \lesssim E_{\rm max, SSDA}$, which may then be smoothly connected to a harder spectrum at high energy $E \gtrsim E_{\rm max, SSDA}$, as predicted by the standard DSA theory. This is the injection scheme that we consider in this study.

If we consider the particle acceleration problem in the infinitely large system, this injection must always happen as long as the scattering rate is finite. The idealized situation is certainly not satisfied in reality. One has to estimate the scattering efficiency to understand whether or not the proposed scheme provides the solution to the injection problem in realistic systems of finite size or finite age.

\subsection{Injection Threshold Energy}
\label{sec:injection-threshold}

Let us now consider particle acceleration in a system of finite spatial extent. In DSA theory, the typical scale size of the particle acceleration region is given by the diffusion length $\ldiff$, which is, in general, energy-dependent. If the diffusion length is comparable to or larger than the system size, no particle acceleration should be expected. Alternatively, one may think of a time-dependent particle acceleration problem. We see from $\tau_{\rm acc} \propto \ldiff \propto (D_{\mu\mu}/\wci)^{-1}$ that the acceleration time becomes longer if the scattering is weaker. Again, particle acceleration will not take place for those particles having a typical acceleration time comparable to or longer than the age of the system. In both cases, the energy dependence of the scattering rate $D_{\mu\mu}$ determines whether or not DSA is capable of accelerating the particles of a given energy. Therefore, the injection from SSDA to DSA requires that the particles escaped out from SSDA (i.e., those with energies $E \sim E_{\rm max, SSDA}$) experience strong enough scattering outside the shock transition layer.

If one considers the standard DSA theory in which the particle scattering outside the shock transition layer is provided by cyclotron resonance with low-frequency magnetostatic turbulence with a power-law spectrum, the scattering rate $D_{\mu\mu}$ will be larger at higher energies. In other words, the resonance condition $k \rg \sim 1$ (where $\rg$ denotes the particle gyroradius) indicates that the resonant wavelength becomes longer as increasing the particle energy. Therefore, higher energy particles will interact with more intense fluctuations at longer wavelengths. Since thermal and slightly supra-thermal electrons have gyroraii much smaller than the typical dissipation length scale of the turbulence, there will be essentially no power available for their scattering. This is the severest obstacle prohibiting the injection of low-energy electrons into DSA. It is clear that the injection threshold energy depends crucially on the assumption of the fluctuation power spectrum both in the upstream and downstream of the shock.

First, we consider pre-existing turbulence in the unshocked medium or that is generated by the shock-accelerated ions. In the former case, a power-law wave spectrum will develop in between the energy injection scale at a long wavelength and the dissipation scale. One may naively estimate that the dissipation scale is determined by the cyclotron damping of thermal ions, which becomes strong at the ion inertial length $k c/\wpi \sim 1$ in a low-beta plasma or the thermal ion gyroradius $k c/\wpi \sim \betai^{-1/2}$ in a high-beta plasma, respectively. Note that $\wpi$ and $\betai$ respectively denote the ion plasma frequency and plasma beta. In either case, it is reasonable to assume that there will be finite wave power for scattering at a wavelength longer than the ion scale. We may apply the same argument for ion-generated turbulence. The primary scale at which specularly reflected ions ($v \sim U_s/\cos \thetabn$) generate large-amplitude fluctuations may be estimated as $k c/\wpi \sim \cos \theta/\MA$. While this appears at a very long wavelength, the development of turbulence will eventually generate fluctuations down to the ion scale. Although we do not exclude the possibility that the spectrum develops into even smaller scales, the integrated power at the electron scale will drop off by orders of magnitude. It is thus reasonable to assume that the turbulence spectrum will have a break at the critical wavenumber $k_{*} c/\wpi \sim {\rm min}(1, \betai^{-1/2})$. Electrons should have sufficiently high energies so that they interact with fluctuations with wavelengths longer than the dissipation scale.

Another possibility is the self-generation of turbulence by the shock-accelerated electrons themselves. The energetic electrons diffuse ahead of the shock may self-generate left-hand polarized \Alfven waves (i.e., \Alfven ion-cyclotron waves) propagating parallel to the ambient magnetic field via the cyclotron resonance. Note that this is exactly the same process that has been considered for the generation of right-hand polarized waves by shock-accelerated ions. The self-generation by the energetic electron streaming will be possible only when the instability overcomes the ion cyclotron damping that becomes strong beyond the critical wavenumber $k_{*}c/\wpi$. We then arrive at the same conclusion that the electrons should have sufficiently high energies for injection such that they can resonantly interact with long-wavelength fluctuations below the critical wavenumber $k_{*} c/\wpi \sim {\rm min}(1, \betai^{-1/2})$.

We should nevertheless also mention that there are possibilities to self-generate (right-hand polarized) higher-frequency whistler waves (either parallel or oblique propagation) by lower-energy electrons \citep{Levinson1992a,Levinson1996,Amano2010}. These models consider that DSA may be activated for low-energy electrons if the self-generation of high-frequency waves is realized. However, the growth rates expected for these instabilities can be very high, and the wave growth and resulting scattering will happen within the shock transition layer. Therefore, we think that these waves will play a role of scattering for SSDA rather than DSA operating outside the shock transition layer. We will discuss this issue in more detail later in Section \ref{sec:discussion}.

\begin{figure}[tbp]
	\centering
	\includegraphics[width=0.45\textwidth]{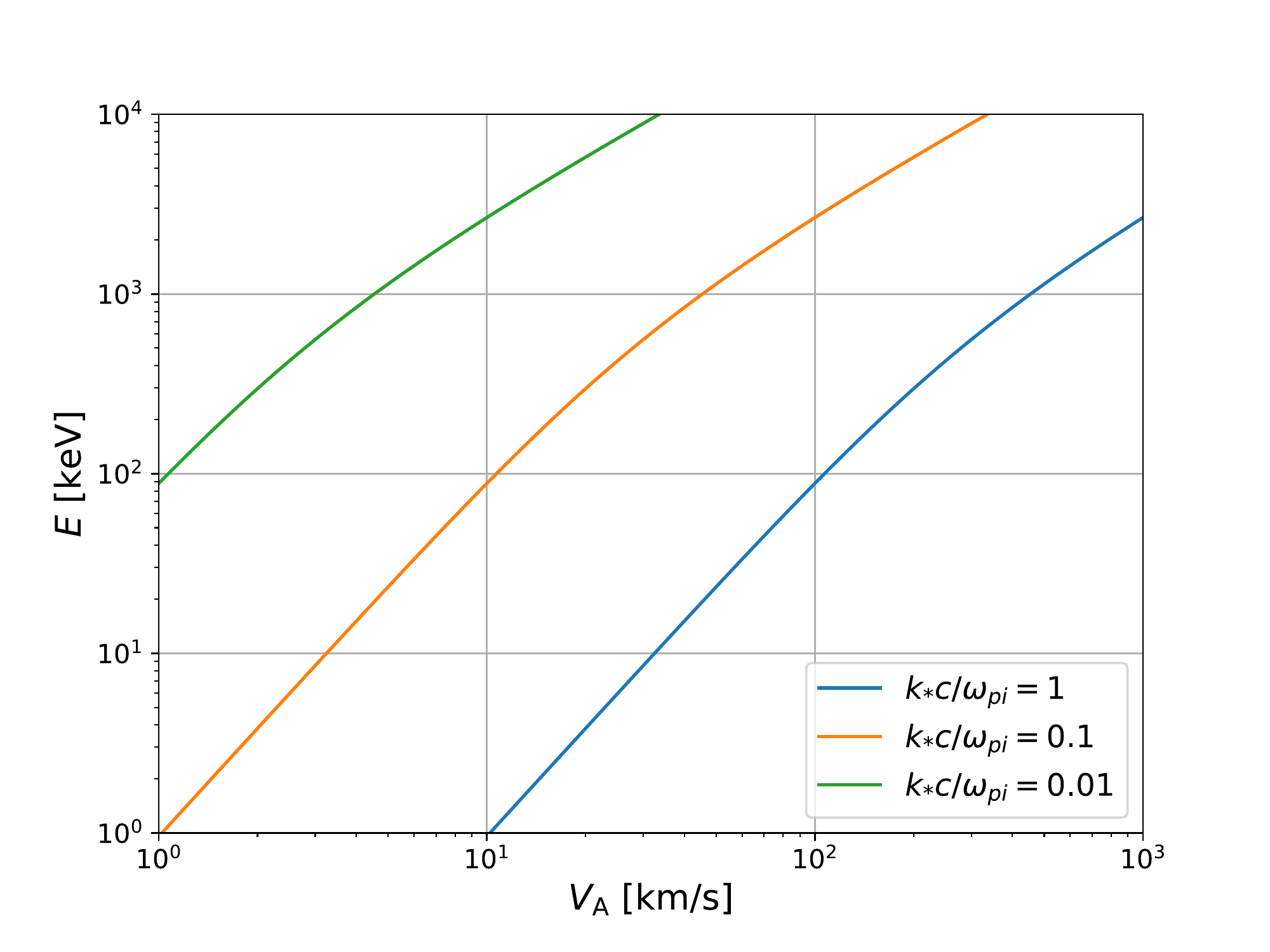}
	\caption{Injection threshold energy as a function of \Alfven speed. The different colors correspond to the different parameters of $k_{*}c/\wpi = 1, 0.1, 0.01$ as indicated in the legend.}
	\label{fig:threshold}
\end{figure}

In summary, we conclude that the threshold energy required for the electron injection is determined by the resonance condition with the wave at the critical wavenumber $k_{*} c/\wpi$. Since the corresponding wave frequency is sufficiently low, we can use the magnetostatic approximation $\omega = k v - \wce/\gamma \approx 0$ (where $\wce$ is the electron cyclotron frequency). We then have the condition
\begin{align}
	\frac{\gamma v}{c} \gtrsim
	\left( \frac{k_{*} c}{\wpi} \right)^{-1}
	\left( \frac{m_i}{m_e} \right)
	\left( \frac{\VA}{c} \right)
	\label{eq:injection-threshold}
\end{align}
as the threshold velocity for electron injection, where $\VA$ is the \Alfven speed. We see that the threshold energy is determined only by the \Alfven speed for a given critical wavenumber $k_{*}$. The corresponding threshold energy as a function of the \Alfven speed is shown in \figref{fig:threshold} for three different parameters of $k_{*}c/\wpi = 1, 0.1, 0.01$. For typical \Alfven speeds of $\VA \sim 100 \, {\rm km/s}$ in the interstellar medium and the solar wind at 1 AU, we may estimate the injection threshold energy as $0.1 \text{--}1 \, {\rm MeV}$ by assuming that the resonant scattering starts to occur below a reasonably small wavenumber $k_{*} c/\wpi \sim 0.1 \text{--} 1$.

It is worth noting that the often-quoted condition that the particle gyroradius is larger than the shock thickness $\rg \gtrsim L_{\rm s}$ is not necessarily relevant for determining the injection threshold. To understand this, let us introduce the mean free path normal to the shock surface $\mfp = v \cos \theta / 2 D_{\mu\mu} = r_g (D_{\mu\mu}/\wce)^{-1} \cos \theta / 2$. For low-energy electrons to which SSDA applies, we have $\mfp/\ldiff = 3 U_{\rm s} / v \cos \theta \ll 1$ because $v \gg U_{\rm s} / \cos \theta$ for the accelerated particles at a non-relativistic shock. As the electron energy increases, the dominant acceleration mechanism will change from SSDA to DSA, which nevertheless happens without violating the condition $\mfp/L_{\rm s} \ll \ldiff/L_{\rm s} \sim 1$. In other words, the particle acceleration for small gyroradius electrons can continue under the diffusion approximation. As the energy increases even further, $\mfp/L_{\rm s} \gtrsim 1$ is reached, for which the electron interaction with the shock is nearly scatter-free. In this case, substantial anisotropy may develop near the shock front because large pitch-angle particles coming from the upstream will be reflected back by the shock via adiabatic SDA, while those coming from the downstream will always be transmitted to the upstream. The strong anisotropy will nevertheless be relaxed a few mean free paths away from the shock, and the overall particle acceleration may be described again by the diffusion approximation. It is well-known that the adiabatic reflection effect is integrated into DSA and will not impose any restrictions on the particle energy. Therefore, if the transition between SSDA and DSA happens under the strong scattering regime such that $L_{\rm s} \gg \mfp \gtrsim r_g$, the diffusion approximation is always applicable. As is clear from the above discussion, the validity of the diffusion approximation will be violated if the HTF shock speed becomes relativistic $U_{\rm s}/\cos \theta \sim c$, which thus requires a separate study. Related discussion on particle acceleration at an oblique shock can be found in, e.g., \citet{Drury1983}.

\subsection{Condition for Electron Injection}
\label{sec:condition-injection}

Let us consider the condition that an electron of a given momentum $p$ can be accelerated by SSDA. We rewrite the maximum energy estimate \eqref{eq:emax-ssda} as the condition required for the \Alfven Mach number in HTF $\MAHTF = \MA / \cos \thetabn$ in the following form:
\begin{align}
	\MAHTF \gtrsim
	\left(  6 \eta \right)^{-1/2}
	\left( \frac{D_{\mu\mu} (p)}{\wce} \right)^{-1/2}
	\left( \frac{p}{\pae} \right),
	\label{eq:ma-cond-p}
\end{align}
where $\pae = m_e V_{\rm A, e} = (m_i m_e)^{1/2} \VA$ is the electron momentum associated with the electron \Alfven speed. Note that we have expressed the momentum dependence of the scattering rate $D_{\mu\mu}(p)$ explicitly. Particle acceleration by SSDA continues as long as this condition is satisfied. Combining this with the injection threshold given by \eqref{eq:injection-threshold}, we estimate the condition for the electron injection:
\begin{align}
	\MAHTF	& \gtrsim
	\left( 6 \eta \right)^{-1/2}
	\left( \frac{D_{\mu\mu}(p)}{\wce} \right)^{-1/2}
	\left( \frac{m_i}{m_e} \right)^{1/2}
	\left( \frac{k_{*} c}{\wpi} \right)^{-1}
	\label{eq:ma-cond-injection}
	\\
	& \approx 55 \,
	\left( \frac{\eta}{1} \right)^{-1/2}
	\left( \frac{D_{\mu\mu}(p)/\wce}{0.1} \right)^{-1/2}
	\left( \frac{k_{*} c / \wpi}{1} \right)^{-1},
	\label{eq:ma-cond-specific}
\end{align}
where the scattering rate $D_{\mu\mu}(p)$ must be evaluated within the shock transition layer and for the threshold momentum corresponding to \eqref{eq:injection-threshold}.

The above conditions for the Mach number $\MAHTF$ suggest that there will be three different regimes of electron acceleration for a given $D_{\mu\mu} (p)$, if it is independent of $\MAHTF$. First, if $\MAHTF$ is too small and is not able to satisfy \eqref{eq:ma-cond-p} for slightly above the thermal momentum, electron acceleration via SSDA will not take place. Second, at higher $\MAHTF$ shocks that satisfy \eqref{eq:ma-cond-p} but not \eqref{eq:ma-cond-injection}, SSDA will be able to accelerate electrons within the shock transition layer. However, the highest-energy electrons escaping out from the system will not experience further energization. Finally, if \eqref{eq:ma-cond-injection} is satisfied, those escaping electrons have sufficiently high energies and will be subject to further particle acceleration via the standard DSA.

In reality, the scattering rate will be dependent on the shock parameters. Indeed, higher $\MAHTF$ shocks may have more efficient scattering rates because of self-generated whistler-mode waves \citep{Levinson1992a,Levinson1996,Amano2010}. Therefore, the problem will be nonlinear and requires in-depth analyses. Nevertheless, it is true that quasi-parallel shocks need to have very high Mach numbers for the electron injection, $\MA \gtrsim 17$ even with the most optimistic Bohm scattering rate $D_{\mu\mu}/\wce \sim 1$. Nearly perpendicular shocks will be more favorable sites for the electron injection as even moderately strong shocks with $\MA \sim 5$ will be sufficient for injection if $\thetabn \sim 85 \degr$ and $D_{\mu\mu}/\wce \sim 0.1$. It is nevertheless important to keep in mind that the required Mach number is dependent on the dissipation scale $k_{*}c/\wpi$ surrounding the shock. The electron injection is regulated by the rate of scattering both inside and outside the shock transition layer.

\section{Numerical Solutions}
\label{sec:numerical-solutions}

\subsection{Setup}
\label{sec:setup}

In this section, we will discuss fully numerical solutions to the diffusion-convection equation \eqref{eq:diffusion-transport} in the steady-state to check the validity of the discussion developed so far. Note that we do not intend to compare the results with specific observations or kinetic simulations. Therefore, we here adopt a simple analytic functional form of the shock profile. We assume that the magnetic field is always lie in the $x$-$z$ plane (i.e., the coplanarity plane) and use the profile for $B_z(x)$ given by
\begin{align}
	B_z (x) =
	\frac{B_{z,2} - B_{z,1}}{2} \tanh \left( \frac{x}{L_s} \right) +
	\frac{B_{z,2} + B_{z,1}}{2}.
\end{align}
The shock compression ratio is given by $r = B_{z,2}/B_{z,1}$. Since a stationary MHD shock satisfies $V_x B_z = {\rm const.}$ across the shock, $V_x(x)$ is readily obtained as:
\begin{align}
	V_x (x) = V_{x,1} \frac{B_{z,1}}{B_z(x)}.
\end{align}
The divergence of the plasma flow is thus given by:
\begin{align}
	\frac{\partial V_x}{\partial x} = - V_x(x) \frac{\partial \ln B_z}{\partial x}.
\end{align}
Note that the conditions $B_x (x) = B \cos \theta = B_1 \cos \thetabn = {\rm const.}$ and $V_z(x) = V  \sin \theta = V_1 \sin \thetabn = {\rm const.}$ always guarantee $\bm{V} \times \bm{B} = 0$.

To represent a finite system size, we introduce the so-called free escaping boundary in the upstream of the shock. In other words, the upstream boundary condition is taken such that $f(x) = 0$ at $x = - L_{\rm FEB}$, indicating that particle acceleration will not take place for those particles with $\ldiff / L_{\rm FEB} \gtrsim 1$. For the downstream boundary condition at $x = +\infty$, we use $\partial f/\partial x = 0$ throughout in this study.

We wish to obtain the solution in the momentum range $p_1 \leq p \leq p_2$. Since $\nabla \cdot \bm{V} \leq 0$ is satisfied throughout the region of interest, there will be no energy loss of the particles. This indicates that the boundary condition at the highest momentum $p = p_2$ is not needed. On the other hand, we have to specify the boundary condition at the lowest momentum $p = p_1$, for which we take the fixed boundary
\begin{align}
	f(x, p_1) = f_{\rm inj}
	\exp \left( - \frac{x^2}{2 \sigma^2} \right),
	\label{eq:boundary-y1}
\end{align}
where $f_{\rm inj}$ is constant. This indicates that there is an injection into the system that is spatially localized at around $x = 0$. Although we always use $\sigma = L_{\rm s}$ in this study, this will not affect the solution as long as we consider the momentum range sufficiently far away from $p_1$. Note that, since the problem is linear, $f_{\rm inj}$ can be chosen arbitrarily, which merely determines the normalization of the numerical solution.

The problem definition will be completed once a model for $\kappa = \kappa(x, p)$ is given. Since the shock transition layer is the region where the main shock dissipation is taking place, the fluctuation level may be substantially different from the regions outside. Therefore, we will consider the solutions in these regions separately, which are then connected with each other through appropriate boundary conditions. More specifically, we divide the entire domain into the following three regions: (1) upstream $-L_{\rm FEB} \leq x \leq X_1$, (2) shock transition layer $X_1 \leq x \leq X_2$, (3) downstream $X_2 \leq x$. By taking $X_1$ and $X_2$ a few times $L_{\rm s}$ away from $x = 0$, we may use the analytic solution for a homogeneous medium both in the upstream and downstream. Taking into account the boundary conditions at $x = -L_{\rm FEB}$ and $x = +\infty$, we have:
\begin{align}
	f(x, p) =
	\begin{cases}
	C^{\UP}(p) \left[
		\exp \left( \dfrac{U_{\rm s} x}{\kappa^{\UP}_{xx}} \right) - 
    \right.
    \\
    \left. \quad\quad\quad
		\exp \left(-\dfrac{U_{\rm s} L_{\rm FEB}}{\kappa^{\UP}_{xx}} \right)
	\right]
	& -L_{\rm FEB} \leq x \leq X_1
	\\
	C^{\DOWN}(p)
    &
	X_2 \leq x
	\end{cases}
	\label{eq:analytic-solution-outside}
\end{align}
for the solutions in the upstream and downstream regions, respectively. Note that $\kappa^{\UP}_{xx} = \kappa^{\UP} (p) \cos^2 \theta_1$ represents the $xx$ component of the momentum-dependent diffusion coefficient in the upstream. Because of the constraint imposed by the boundary condition at $x = +\infty$, the problem is independent of the downstream diffusion coefficient. We will denote the diffusion coefficient inside the shock transition layer by $\kappa$ without the superscript ${}^{\UP}$, which may or may not be different from $\kappa^{\UP}$. The momentum-dependent unknown constants $C^{\UP}(p)$, $C^{\DOWN}(p)$ must be determined by connecting these solutions to the one inside the shock transition layer.

In general, it is difficult to obtain a simple analytic solution in the region $X_1 \leq x \leq X_2$. Therefore, we expand the solution using orthogonal basis functions and use the pseudospectral method. We demand the continuity of the density $f$ and the flux $\kappa_{xx} \partial f/ \partial x$ as the boundary condition at $x = X_1$ and $x = X_2$ for connecting the internal and external solutions. Note that the first-order derivative $\partial f/\partial x$ may be discontinuous if the diffusion coefficient changes discontinuously across the boundary. Ultimately, the problem is reduced to solving a large system of linear equations. More detailed descriptions of numerical techniques may be found in Appendix \ref{sec:appendix-method}.

The following parameters will always be fixed for the numerical solutions discussed below: $r = 4$, $B_1 = 1$, $U_{\rm s} = 1$, $L_{\rm s} = 1$, $X_2 = -X_1 = 4 L_{\rm s}$. The orders of expansion for the orthogonal basis in $x$ and $y$ directions will be denoted by $N_x$ and $N_y$, respectively.

\subsection{Constant Diffusion Coefficient}
\label{sec:constant}

\begin{figure}[htp]
	\centering
	\includegraphics[width=0.45\textwidth]{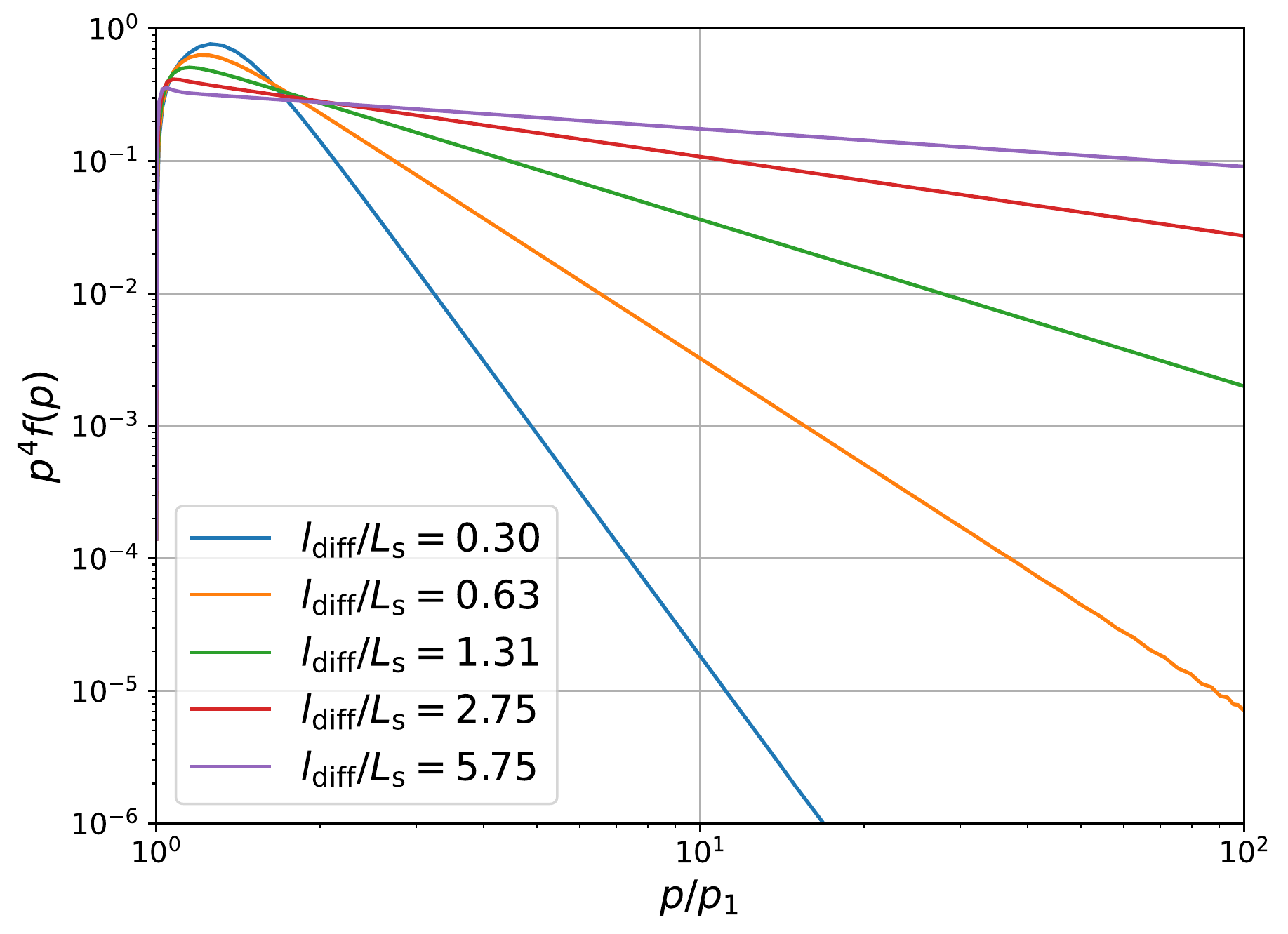}
	\caption{Downstream momentum spectra for $\thetabn = 60\degr$ obtained with constant diffusion coefficients. The numerical solutions obtained with $\ldiff/L_{\rm s} = 0.30, 0.63, 1.31, 2.75, 5.75$ are shown with different colors.}
	\label{fig:ldiff_dependence1}
\end{figure}

\begin{figure}[tbp]
	\centering
	\includegraphics[width=0.45\textwidth]{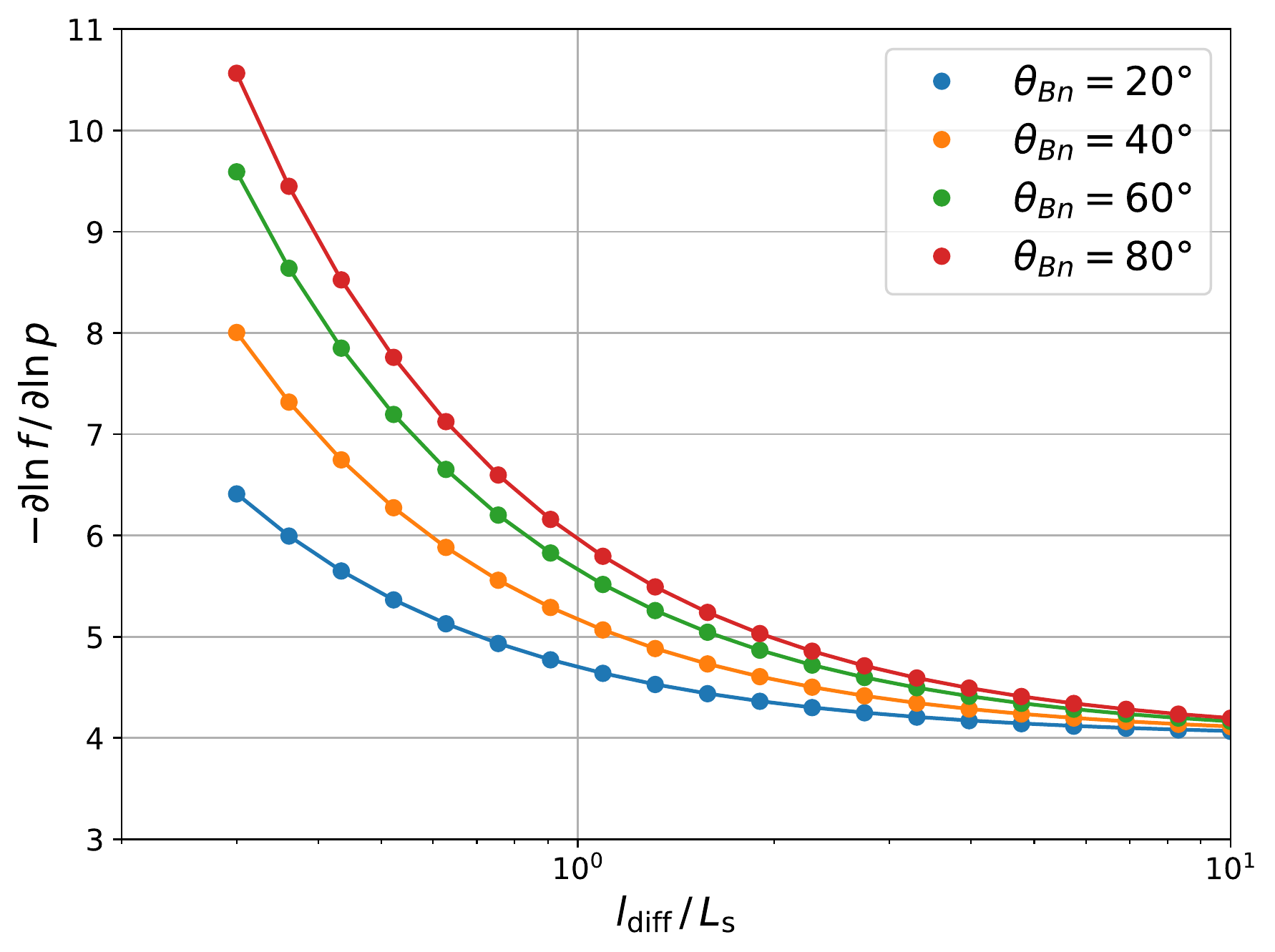}
	\caption{Dependence of spectral index on $\thetabn$ and $\ldiff/L_{\rm s}$. The numerical solutions obtained with $\thetabn = 20\degr, 40\degr, 60\degr, 80\degr$ are shown with different colors. The filled circles indicate the indices obtained by direct numerical derivatives $-\partial \ln f/\partial \ln p$ evaluated at $p/p_1 = 5$. The solid lines are obtained by integrating \eqref{eq:general-index-formula} using the spatial profiles $f(x)$ and $V_x(x)$.}
	\label{fig:ldiff_dependence2}
\end{figure}

We first consider the simplest case in which the diffusion coefficient is constant $\kappa = \kappa^{\UP} = \kappa_0$ both in space and momentum. Parameters used for the results discussed in this subsections are as follows: $N_x = 32$, $N_y = 64$, $p_2/p_1 = 10^{2}$, $L_{\rm FEB}/L_{\rm s} = 10^{5}$. Note that the results do not depend of the free escaping boundary as it is taken sufficiently far away from the shock.

\figref{fig:ldiff_dependence1} shows the downstream momentum spectra for $\thetabn = 60 \degr$ obtained with several different values of $\ldiff/L_{\rm s}$. Note that we always show the spectrum multiplied by $p^4$ so that the standard DSA limit will be represented as a flat spectrum. We confirm that, as soon as the momentum becomes a few times larger than the injection momentum, a power-law spectrum develops. The spectrum is much steeper for $\ldiff/L_{\rm s} \lesssim 1$ and approaches gradually to $q = 4$ as increasing $\ldiff/L_{\rm s}$, which is consistent with the theoretical prediction. We note that a similar conclusion was also obtained with an analytic approach by \citet{Drury1982}.

The dependence on $\ldiff/L_{\rm s}$ is more clearly seen in \figref{fig:ldiff_dependence2}. The spectral indices evaluated directly from the numerical solutions by calculating numerical derivatives $\partial \ln f/\partial \ln p$ at $p/p_1 = 5$ are shown with the filled circles. We have also calculated the spectral indices by integrating \eqref{eq:general-index-formula} using the numerical solutions, which are shown with the solid lines. It is seen that they both agree very well, indicating that \eqref{eq:general-index-formula} is indeed useful to calculate a theoretical spectrum from known profiles of $f(x)$ and $V_x(x)$.

As is clearly seen, the spectral index approaches to the DSA prediction at $\ldiff/L_{\rm s} \gg 1$ and steepens substantially for $\ldiff/L_{\rm s} \lesssim 1$. While this trend itself is always true, the actual slope is dependent on $\thetabn$ as well. This is because the gradual kink in the magnetic field line changes the diffusive particle flux. Although the particle diffusion always occurs along the local field line, the decrease in $\theta$ from the downstream to the upstream side increases the particle flux toward the negative $x$ (i.e., upstream) direction when the magnetic curvature is non-negligible. This makes the effective diffusion length longer at small $\thetabn$. Since $\theta \approx \thetabn \approx {\rm const}$ at quasi-perpendicular shocks, this effect is negligible and the steeper spectra will result for larger $\thetabn$.

\subsection{Bohm-like Diffusion Coefficient}
\label{sec:bohm-like}

We now consider the diffusion coefficient given by $\kappa (p) = \kappa^{\UP} (p) = \kappa_1 \left( p/p_1 \right)$, where $\kappa_1$ is constant. It is often used as a phenomenological diffusion model for CR transport as it implies that $\mfp/r_g = {\rm const.}$ for relativistic particles. We here employ the model simply to show qualitative characteristics expected for a diffusion coefficient monotonically increasing with momentum. We use $N_x = 32$, $N_y = 128$, $p_2/p_1 = 10^{4}$, $\thetabn = 80 \degr$ and $\kappa_1 = 10$. This choice makes the diffusion length for the lowest momentum $p = p_1$ smaller than the shock thickness $\ldiff/L_{\rm s} = \kappa_1 \cos^2 \thetabn / (U_{\rm s} L_{\rm s}) \approx 0.30$. The diffusion length increases with momentum and eventually becomes much larger than the thickness $\ldiff/L_{\rm s} \gg 1$. Therefore, we expect that SSDA should be dominant for low-energy particles and the acceleration regime gradually shifts to DSA at higher energies.

\begin{figure}[tbp]
	\centering
	\includegraphics[width=0.45\textwidth]{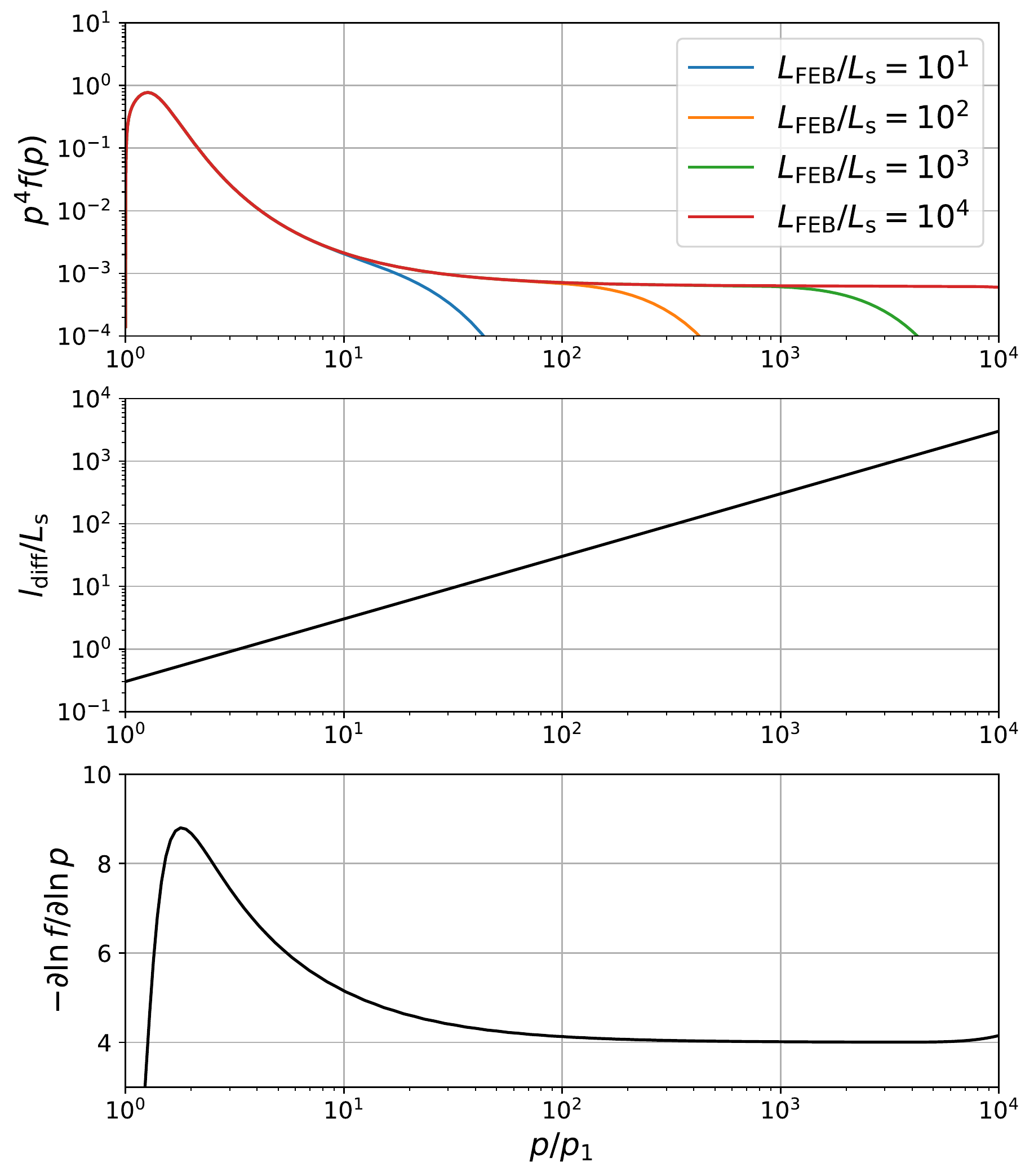}
	\caption{Numerical solutions obtained with Bohm-like diffusion coefficient model. The momentum spectra obtained with $L_{\rm FEB}/L_{\rm s} = 10^{1}, 10^{2} 10^{3}, 10^{4}$ (top), $\ldiff/L_{\rm s}$ for $L_{\rm FEB}/L_{\rm s} = 10^{4}$ (middle), local spectral slope $-\partial \ln f/\partial \ln p$ for $L_{\rm FEB}/L_{\rm s} = 10^{6}$ (bottom) are shown.}
	\label{fig:bohmlike_kappa}
\end{figure}

The top panel of \figref{fig:bohmlike_kappa} shows the downstream spectra obtained with $L_{\rm FEB}/L_{\rm s} = 10^{1}, 10^{2}, 10^{3}, 10^{4}$. The middle and bottom panels respectively show $\ldiff/L_{\rm s}$ as a function of momentum and the local spectral slope for $L_{\rm FEB}/L_{\rm s} = 10^{4}$ evaluated by direct numerical derivative. We find that the spectrum at low energy $p/p_1 \lesssim 10$ is very steep. It is clear that the spectrum is not a pure power law and has a concave shape. This is qualitatively consistent with the increasing trend of $\ldiff/L_{\rm s}$ as a function of momentum. We see from the middle panel that $\ldiff/L_{\rm s} \gtrsim 10$ is reached for $p/p_1 \gtrsim 30$, where the slope becomes close to the canonical DSA prediction as seen in the bottom panel. The high-energy part of the spectrum will be cut off if the particles diffusing ahead of the shock reach the free escaping boundary $\ldiff/L_{\rm s} \sim 1$. Comparison between the top and middle panels confirms that the idea is consistent with the numerical solutions.

We think that the spectrum shown in \figref{fig:bohmlike_kappa} qualitatively represents the entire history of particle acceleration: injection by SSDA into DSA at low energy, further acceleration by DSA, and escape of the highest energy particles from the system resulting in the high-energy cutoff. With a more complicated but ``realistic'' momentum- and space-dependent diffusion model, we investigate the possibility of injection more quantitatively in the next subsection.

\subsection{``Realistic'' Diffusion Coefficient}
\label{sec:realistic}

\begin{figure}[htbp]
	\centering
	\includegraphics[width=0.45\textwidth]{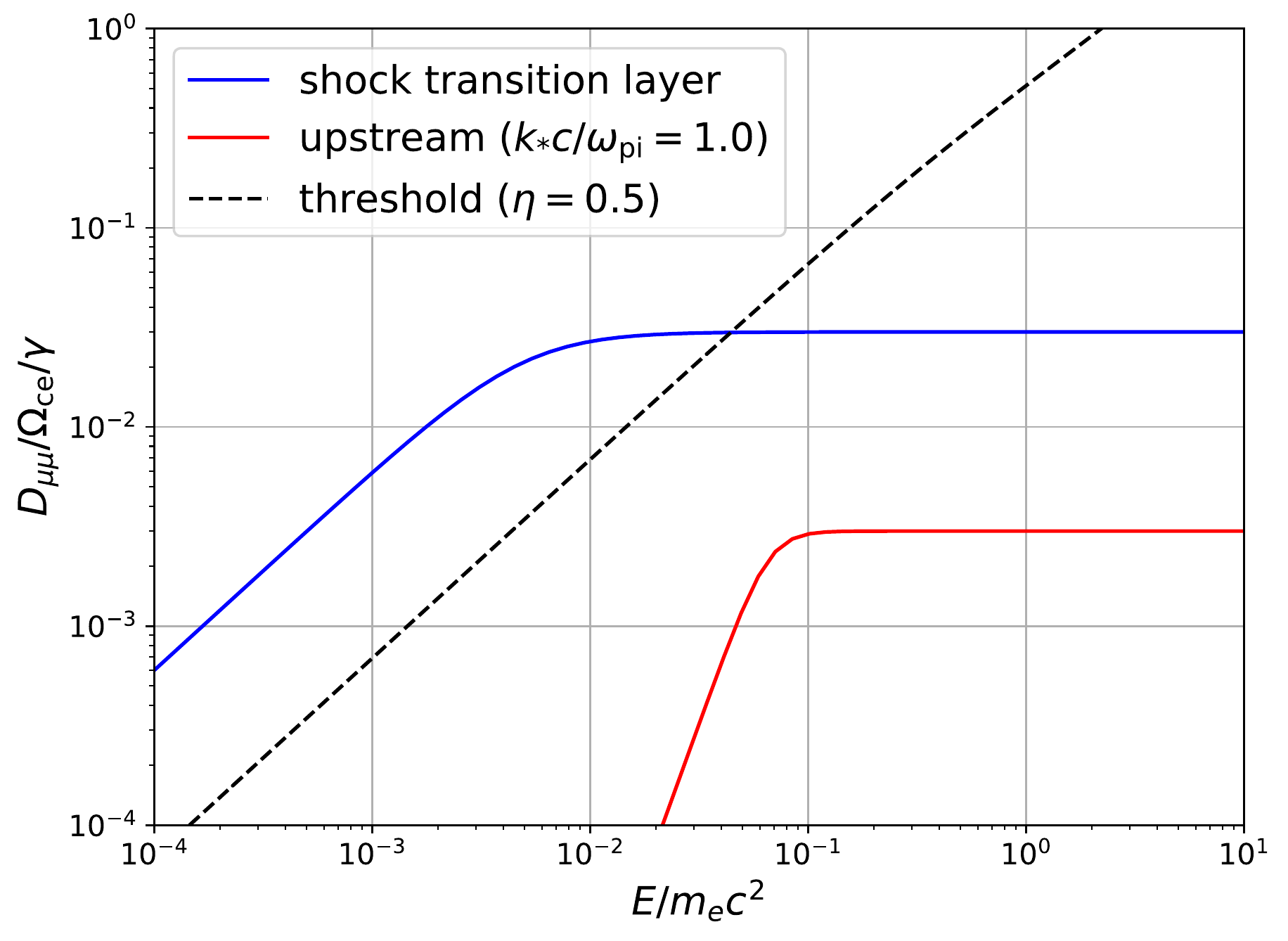}
	\caption{Model of scattering rate as a function of energy. The blue line indicates the scattering rate within the shock transition layer, while the red line is for the upstream region. The black dashed line represents the theoretical threshold given by $\ldiff/L_{\rm s} = 1$, indicating that SSDA is effective only when the scattering rate is larger than the threshold.}
	\label{fig:scattering_rate}
\end{figure}

\begin{figure}[tbp]
	\centering
	\includegraphics[width=0.45\textwidth]{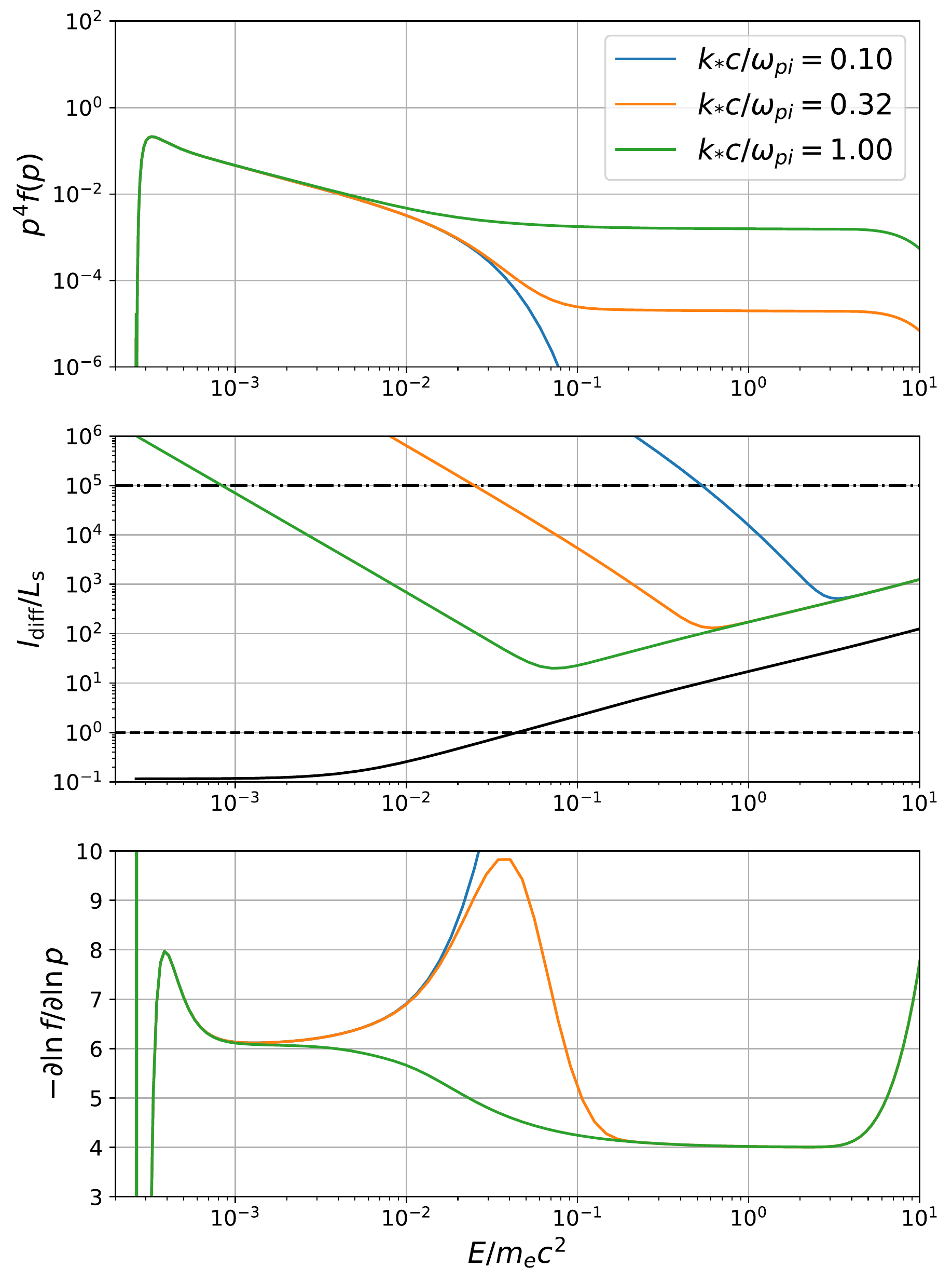}
	\caption{Numerical solutions obtained with $\thetabn = 85 \degr$ and $k_{*}c/\wpi = 0.10, 0.32, 1.0$. The energy spectra (top), $\ldiff/L_{\rm s}$ (middle), and local spectral slopes $-\partial \ln f/\partial \ln p$ (bottom) are shown. The different colors indicate different $k_{*}c/\wpi$ as shown in the legend on the top panel, except for the black solid line in the middle panel showing $\ldiff/L_{\rm s}$ within the shock transition layer that is independent of $k_{*} c/\wpi$. The horizontal dashed and dash-dotted lines in the middle panel correspond respectively to $\ldiff/L_{\rm s} = 1$ and $\ldiff/L_{\rm FEB} = 1$ for reference.}
	\label{fig:realistic_ks}
\end{figure}

\begin{figure}[tbp]
	\centering
	\includegraphics[width=0.45\textwidth]{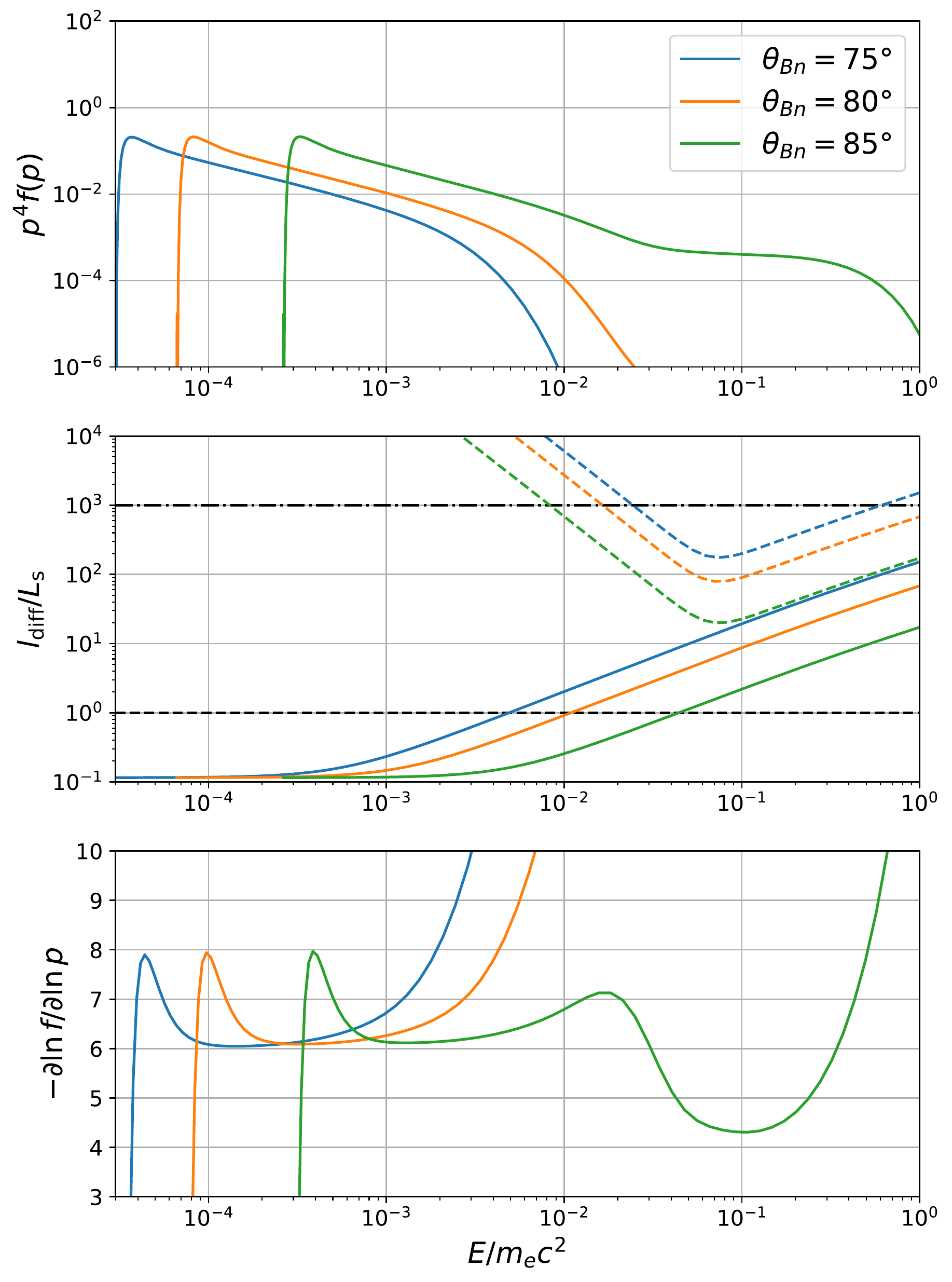}
	\caption{Numerical solutions obtained with $\thetabn = 75 \degr, 80 \degr, 85 \degr$ and $k_{*}c/\wpi = 1.0$. The energy spectra (top), $\ldiff/L_{\rm s}$ (middle), and local spectral slopes $-\partial \ln f/\partial \ln p$ (bottom) are shown. The different colors indicate different $\thetabn$ as shown in the legend on the top panel. The solid and dashed lines in the middle panel indicate $\ldiff/L_{\rm s}$ for the shock transition layer and the upstream region, respectively. The horizontal dashed and dash-dotted lines in the middle panel correspond respectively to $\ldiff/L_{\rm s} = 1$ and $\ldiff/L_{\rm FEB} = 1$ for reference.}
	\label{fig:realistic_theta}
\end{figure}

\begin{figure}[tbp]
	\centering
	\includegraphics[width=0.45\textwidth]{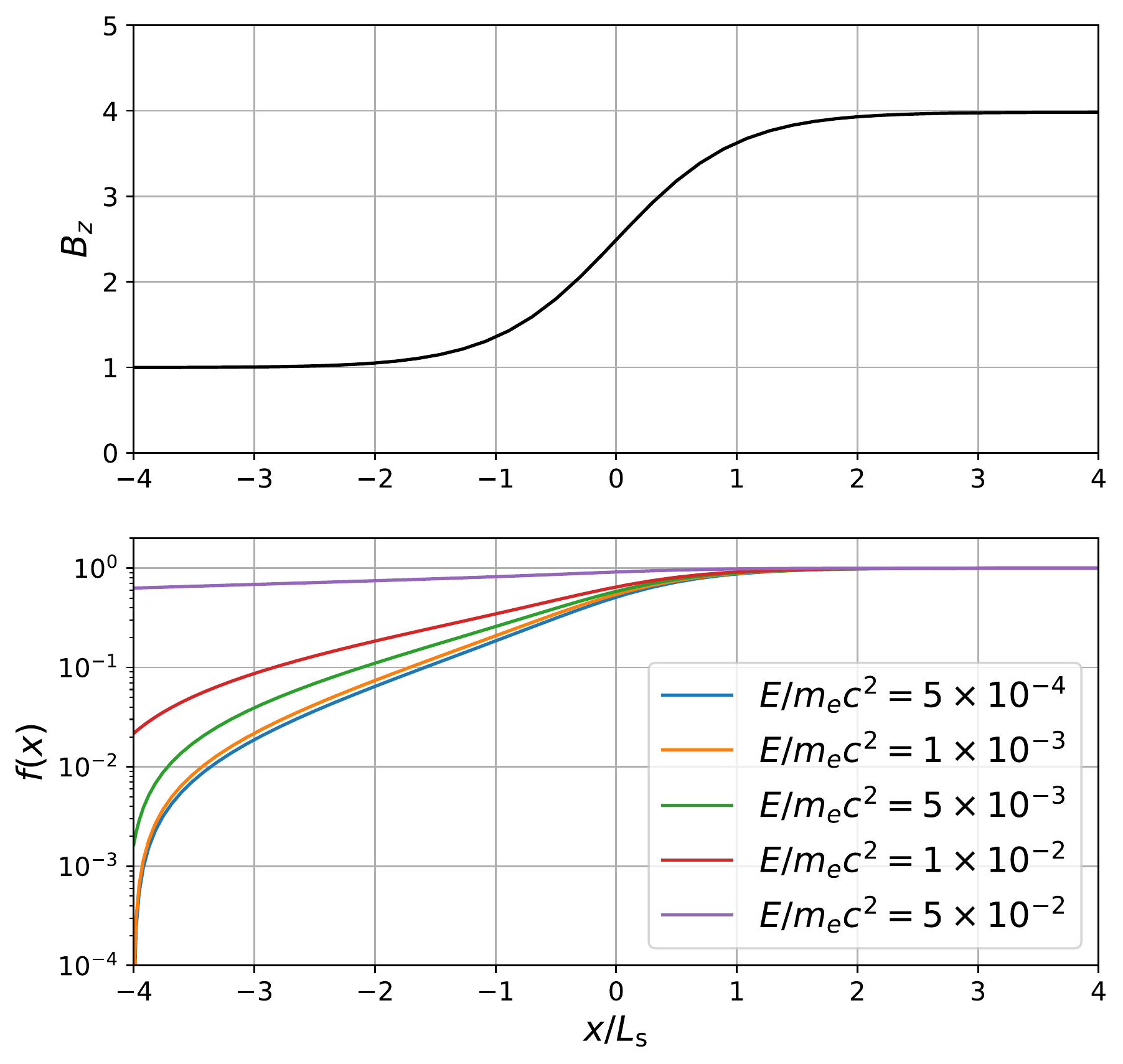}
	\caption{Spatial profiles of distribution function. The numerical solution is obtained with $\thetabn = 85\degr$. The spatial profile of $B_z(x)$  (top), and the distribution function $f(x)$ at various energies (bottom) are shown. The distribution function is normalized to the downstream value for each energy.}
	\label{fig:realistic_spatial1}
\end{figure}

\begin{figure}[tbp]
	\centering
	\includegraphics[width=0.45\textwidth]{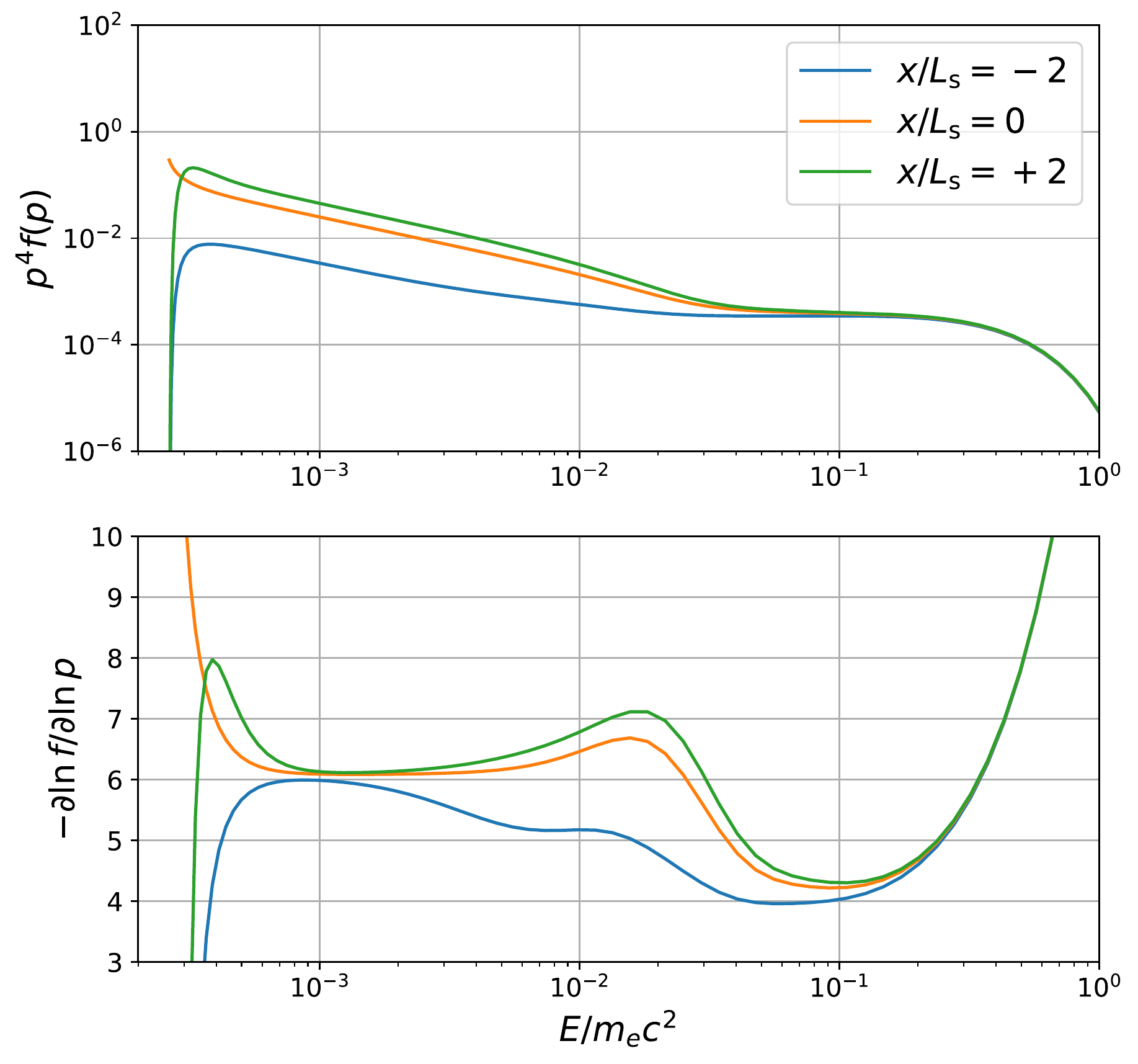}
	\caption{Energy spectra at different spatial positions. The numerical solution is obtained with $\thetabn = 85\degr$. The energy spectra (top) and local spectral slopes (bottom) are shown. The different colors indicate the spectra at different positions $x/L_{\rm s} = -2, 0, +2$ shown in the legend on the top panel.}
	\label{fig:realistic_spatial2}
\end{figure}

For the numerical solutions discussed so far, we have not specified either a physical
scale length for the shock thickness or a time scale of particle pitch-angle scattering. For a more quantitative discussion of electron injection, we here introduce specific physical scales based on available knowledge of the collisionless shock dynamics.

It is physically more transparent to define the scattering rate normalized to the gyrofrequency $\wce/\gamma$ (including the relativistic effect) and convert it to the diffusion coefficient. We thus introduce the following functional form for the pitch-angle scattering rate:
\begin{align}
	\frac{D_{\mu\mu} (p)}{\wce/\gamma} = \tilde{D}_{\mu\mu,0}
	\left[
		1 + \left( \frac{p}{p_*} \right)^{-2 \nu}
	\right]^{-1/2},
	\label{eq:scattering-rate}
\end{align}
where $\tilde{D}_{\mu\mu,0}$ is a dimensionless constant. For non-relativistic energies and with $\nu > 0$, it smoothly connects $D_{\mu\mu}(p) \propto p^{\nu}$ for $p \ll p_*$ and $D_{\mu\mu} (p) \approx {\rm const.}$ for $p \gg p_*$. For relativistic energies, it approaches $D_{\mu\mu} (p) \propto p^{-1}$, which gives a constant scattering rate when normalized to the relativistic gyrofrequency $\wce/\gamma$. It is reasonable to consider the scattering rate monotonically increasing with momentum because higher-energy particles can resonantly interact with lower-frequency and larger amplitude waves. On the other hand, it is natural to assume that the normalized scattering rate saturates to a constant at high energies because it is likely to be limited by the Bohm scattering rate $D_{\mu\mu} \sim \wce/\gamma$. \eqref{eq:scattering-rate} qualitatively represents such a momentum dependence.

Since it is not an easy task to obtain an accurate and physically grounded analytic model for the scattering rate within the shock transition layer, we assume a phenomenological model based on recent in-situ observations \citep{Amano2020}. The particle intensity profiles measured at Earth's bow shock indicate that $D_{\mu\mu} \propto p^{2}$ at energies below a few keV, while it saturates to a constant at higher energies. This motivates us to choose $\nu = 2$, $\tilde{D}_{\mu\mu,0} = 3 \times 10^{-2}$, $p_{*}/(m_e c) = 1 \times 10^{-1}$ (which  corresponds to $\sim 2.5$ keV) as fiducial parameters. It is interesting to note that $D_{\mu\mu} \propto p^2$ at lower energies indicates that the diffusion length is energy independent. Therefore, we expect that the spectrum in this energy range is described by a single power law, which is consistent with the measured particle spectrum. The saturation of $D_{\mu\mu}$ at higher energies indicates that the diffusion length monotonically increases with energy, leading to the escape of higher-energy particles from the shock transition layer.

The particles with diffusion lengths longer than the shock thickness primarily interact with turbulence in the upstream and downstream regions. As we have discussed in Section \ref{sec:injection-threshold}, the fluctuation power available for particle scattering will be very small at short wavelength $k \gtrsim k_{*}$. Therefore, we assume that the scattering rate is finite only for those particles with momenta larger than the threshold $p \gtrsim p^{\UP}_{*}$ but is essentially negligible otherwise. Note that we denote quantities defined in the upstream by the superscript ${}^{\UP}$. For simplicity, we use the same functional form \eqref{eq:scattering-rate} with $\nu^{\UP} = 6$, $\tilde{D}^{\UP}_{\mu\mu} = 3 \times 10^{-3}$, and $p^{\UP}_{*}/(m_e c) = (k_{*} c / \wpi)^{-1} (m_i/m_e)$, where the last relationship between $p^{\UP}_{*}$ and $k_{*}$ comes from \eqref{eq:injection-threshold}.

In the following, unless otherwise noted, we use the following parameters representative of a moderately high Mach number and quasi-perpendicular Earth's bow shock: $\MA = 10$, $\thetabn = 85 \degr$, $p_1 = m_e U_{\rm s} / \cos \thetabn$, $p_2/p_1 = 10^{3}$, $\VA/c = 2 \times 10^{-4}$, $\eta = 0.5$. Note that we determine the lower bound of momentum by assuming that particle acceleration starts from the shock speed in HTF $v \sim U_{\rm s}/\cos \thetabn$. \figref{fig:scattering_rate} shows the scattering rates as functions of energy for the shock transition layer (blue) and the upstream (red) as defined above. The threshold scattering rate defined by $\ldiff/L_{\rm s} = 1$ is also shown with the black dashed line. Note that we have assumed $k_{*} c / \wpi = 1$. In this case, since particles with energies $E/m_e c^2 \lesssim 5 \times 10^{-2}$ (i.e., $\lesssim 25$ keV) have the diffusion length smaller than the shock thickness, their major acceleration mechanism will be SSDA. In particular, a pure power-law spectrum is expected for $E/m_e c^2 \lesssim 5 \times 10^{-3}$ (i.e., $\lesssim 5$ keV) for which $D_{\mu\mu} \propto p^2$ ($\ldiff / L_{\rm s} \approx {\rm const.}$) As the maximum energy expected for SSDA $E_{\rm max, SSDA}/m_e c^2 \sim 5 \times 10^{-2}$ is comparable to the threshold energy associated with $p^{\UP}_{*}$ (which corresponds to $E/m_e c^2 \sim 10^{-1}$), higher-energy particles escaping out from the shock transition layer will be injected into the particle acceleration cycle through DSA.

\figref{fig:realistic_ks} confirms the electron injection scenario, in which numerical solutions for $k_{*} c/\wpi = 0.1, 0.32, 1.0$ are shown with $L_{\rm FEB}/L_{\rm s} = 10^{5}$. The spectra on the top panel clearly show that the low-energy steeper power-law spectrum $E/m_e c^2 \lesssim 10^{-2}$ is smoothly connected to the flat DSA spectrum at high energies for $k_{*} c/\wpi = 1.0$. As decreasing $k_{*} c/\wpi$, however, the injection becomes less efficient. The middle panel shows $\ldiff/L_{\rm s}$ for the upstream and the shock transition layer as functions of energy. We see that $\ldiff/L_{\rm FEB} \ll 1$ for $k_{*} c/\wpi = 1.0$ at the energy where $\ldiff/L_{\rm s} \sim 1$. Therefore, DSA is applicable at this energy. On the other hand, since $\ldiff/L_{\rm FEB} \sim 1$ for $k_{*} c/\wpi = 0.32$ at the same energy, a substantial fraction of particles will be lost from the system. This explains the partial cutoff before the spectrum becomes harder. With $k_{*} c/\wpi = 0.1$, the cutoff is almost complete as expected from $\ldiff/L_{\rm FEB} \gg 1$. In the bottom panel, local spectral slopes calculated by numerical derivative are shown as functions of energy. In all three cases, the low-energy spectral indices are given by $q \sim 6$. At sufficiently high energies, on the other hand, the spectral indices (for $k_{*} c/\wpi = 0.32, 1.0$) are consistent with $q = 4$ as predicted by DSA. Note that the highest-energy cutoff is due to the free escaping boundary.

We have seen that the upstream turbulence property as characterized by $k_{*}$ is important to control the electron injection. Similarly, the capability of electron injection is affected strongly by $\MA$ and $\thetabn$ as predicted by \eqref{eq:ma-cond-injection}. To see this, the numerical solutions obtained with $\thetabn = 75\degr, 80\degr, 85\degr$ and a fixed value of $L_{\rm FEB}/L_{\rm s} = 10^{3}$ are compared in \figref{fig:realistic_theta} in a similar format. Note that we have employed the scaling $p_{*} \propto m_e U_{\rm s}/\cos \thetabn$, which is the same as $p_1$ and $p_2$, assuming that the velocity scales with the shock speed in HTF $U_{\rm s}/\cos \thetabn$. On the other hand, $p^{\UP}_{*}$ does not depend on the shock parameters and a fixed value of $k_{*} c/\wpi = 1.0$ is used for all three cases. The low-energy part of spectra in three cases are very similar but shifted in energy because of the energy scaling $\propto 1/\cos^2 \thetabn$. As increasing the shock obliquity, the maximum energy of SSDA increases and eventually reaches the injection threshold. This results in the flat spectrum above $E/m_e c^2 \sim 3 \times 10^{-2}$ seen only for $\thetabn = 85 \degr$. As in the case of \figref{fig:realistic_ks}, the characteristics of the spectra can be reasonably understood by looking at the energy dependence of $\ldiff/L_{\rm s}$ shown in the middle panel. More specifically, $\ldiff/L_{\rm FEB} \ll 1$ is satisifed only for $\thetabn = 85 \degr$ at the theoretical maximum energy for SSDA determined by $\ldiff/L_{\rm s} \sim 1$.

The result shown here is consistent with the discussion presented in Section \ref{sec:condition-injection}, in particular, \eqref{eq:ma-cond-injection}. We note that the model for the scattering rate adopted here has a relatively smooth transition below and above $p^{\UP}_{*}$. Also, the choice of $\nu^{\UP} = 6$ makes the condition for electron injection more or less sensitive to $L_{\rm FEB}$. The sensitivity on $L_{\rm FEB}$ will be eliminated if one chooses a model that has a more abrupt transition of energy dependences across $p^{\UP}_{*}$. In any case, the efficiency of particle acceleration within the shock transition layer alone is not able to determine the injection condition. Depending on the efficiency and the energy dependence of the scattering rate outside the shock transition layer, the injection may or may not be achieved in a system of a given scale size $L_{\rm FEB}$.

Although we have concentrated only on the downstream spectrum so far, it is also important to look at the spatial dependence of the spectrum for comparison with observations and kinetic simulations. \figref{fig:realistic_spatial1} shows the profiles of magnetic field $B_z(x)$, as well as the distribution functions $f(x)$ at specific energies for the solution obtained with $\thetabn = 85 \degr$. Note that $f(x)$ is normalized by the downstream value for each energy. We see that the profiles at low energies are nearly identical with each other, which is a natural consequence of $\ldiff/L_{\rm s} \approx {\rm const}$. The characteristics length scale becomes larger as increasing energy. The profile at the highest energy is nearly flat, meaning that $\ldiff/L_{\rm s} \gg 1$. This is consistent with the fact that the particle acceleration at this energy is essentially the same as DSA at a discontinuous shock. Recall that the solution within the shock transition layer shown in \figref{fig:realistic_spatial1} is connected to the external (analytic) solutions with the boundary conditions. Therefore, the abrupt drop down in density at $x/L_{\rm s} = -4$ seen at low energies is not a boundary effect but due to the physical escape beyond the free escaping boundary.

\figref{fig:realistic_spatial2} shows the energy spectra and spectral slopes measured at different spatial positions $x/L_{\rm s} = -2, 0, +2$. At low energy $E/m_e c^2 \lesssim 3 \times 10^{-3}$, the spectral slopes are nearly independent of $x$ while the density becomes substantially larger as approaching the downstream. This is again due to the constant $\ldiff/L_{\rm s}$. The immediate upstream spectrum beyond this energy is harder than the downstream because of the monotonically increasing diffusion length as a function of energy. The spatial dependence of the spectrum provides yet another clue to test the theory against observations and kinetic simulations.

\section{Discussion}
\label{sec:discussion}

In this paper, we have presented a detailed analysis of SSDA and discussed its relation to DSA. We have shown that both SSDA and the conventional DSA are special cases of particle acceleration at an oblique shock of finite thickness under the diffusion approximation. More specifically, it is the ratio between the diffusion length to the shock thickness $\ldiff/L_{\rm s}$ that differentiates the two particle acceleration regimes. If the ratio is much larger than unity $\ldiff/L_{\rm s} \gg 1$, the shock should be seen as a discontinuity by the accelerated particles. In this case, the solution naturally reduces to the conventional DSA. On the other hand, if the ratio becomes of order unity or smaller $\ldiff/L_{\rm s} \lesssim 1$, the finite extent of the shock structure has a direct influence on the particle acceleration. The resulting spectrum becomes, in general, steeper than the DSA prediction and is not necessarily described by a pure power law. A power-law spectrum is realized only when the diffusion length is independent of energy.

If the diffusion length increases with energy, at some point, it becomes larger than the shock thickness. Beyond this energy, SSDA no longer applies, and the particles escaping out from the shock may be accelerated further via the standard DSA instead. In other words, the particles that reach the maximum energy achievable through SSDA may be injected into DSA operating in a much larger spatial extent. In this scenario, the pre-acceleration by SSDA and the main acceleration by DSA are both consistently described by the unified model.

The injection into DSA is, however, not always possible. In short, the pitch-angle scattering efficiency both inside and outside the shock transition layer is the key for the injection. The scattering inside the shock transition layer determines the maximum energy of SSDA, while that outside controls the injection threshold energy beyond which DSA reasonably operates. An efficient injection requires that the former is comparable to or larger than the latter. For a given efficiency of pitch-angle scattering, the injection is controlled mainly by the parameter $\MAHTF = \MA / \cos \thetabn$. Although the injection threshold energy depends crucially on the dissipation scale of turbulence, we estimate it to be roughly around $0.1\text{--}1$ MeV in typical interstellar or interplanetary plasma conditions. We emphasize that it is much smaller than the energy at which the electron gyroradius becomes comparable to the shock thickness, which is often assumed as the injection threshold. We have argued that, as long as the diffusion approximation is appropriate, the acceleration of electrons should be reasonably described by the present model even when the gyroradius is smaller than the shock thickness.

Admittedly, we have made a number of simplifying assumptions in trying to single out the essential ingredient of electron injection. There thus remains room for further improvement to make more quantitative predictions. Nevertheless, the theoretical framework developed in this paper provides a significant step forward for resolving the issue of electron injection. In the following, we discuss the remaining issues and future perspectives.

\subsection{Self-Generation of Plasma Waves}
\label{sec:self-generation}

In the proposed scenario of electron injection, the pitch-angle scattering over a wide range of energy plays a crucial role. The ``realistic'' model that we discussed in Section \ref{sec:realistic} adopts a phenomenological model based on in-situ measurements of a single bow shock crossing event. The universality of the result is thus not guaranteed at present. In particular, since we know that the collisionless shock dynamics is largely controlled by parameters such as $\MA$ and $\thetabn$, it is quite likely that the wave intensity may also vary substantially depending on these parameters. We should consider the generation of plasma waves over a broad range of frequencies corresponding to the broad range of particle energy. Based on in-situ measurements, the waves of particular interest may be subdivided into three characteristic frequencies: (1) high-frequency quasi-parallel whistlers $0.1 \lesssim \omega/\wce \lesssim 0.5$ \citep[e.g.,][]{Zhang1999,Hull2012,Oka2017}, (2) low-frequency oblique whistlers $\omega \sim \omega_{\rm LH}$, where $\omega_{\rm LH}$ is the lower-hybrid frequency \citep{Hull2012,Oka2019,Hull2020}, (3) Alfv\'en-ion-cyclotron (AIC) waves or the rippling mode $\omega \lesssim \wci$ \citep[e.g.,][]{Winske1988,Johlander2016}. More details on these wave characteristics will be discussed in Appendix \ref{sec:multiscale-wave}.

Among these, the high-frequency whistler waves are the key ingredient for triggering electron acceleration in the first place. In particular, because the wave intensity is much smaller at higher frequency in general, the condition $\ldiff/L_{\rm s} \lesssim 1$ is more difficult to satisfy at low energy than high energy. Interestingly, the spatial profiles of nonthermal electrons indicate that $\ldiff/L_{\rm s} \approx {\rm const.}$ is satisfied for the energy range where a power-law spectrum is observed. This is quite consistent with the theoretical prediction. It might thus be possible to speculate that the generation of both high- and low-frequency whistler waves and the resulting electron confinement occurs in a self-sustaining manner. In other words, if the weakly scattered accelerated electrons tend to escape upstream, the electron streaming becomes more intense and may destabilize waves that scatter themselves to reduce the diffusion length to be comparable to the shock thickness. Unless there exists such a self-sustaining mechanism, it is hard to believe that $\ldiff/L_{\rm s} \approx {\rm const.}$ is satisfied merely by chance.

The self-generation of waves by the back-streaming low-energy electrons that can scatter themselves has been discussed theoretically. \citet{Levinson1992a,Levinson1996} proposed that upstream-directed oblique and low-frequency whistler-mode waves may be generated by low-energy electrons that obey the diffusion-convection equation. The instability condition for an oblique shock may be written as $\MAHTF \gtrsim (m_i/m_e)^{1/2} \betae^{-1/2}$ where $\betae$ is the electron plasma beta. Note that the $\betae$ dependence appears if the thermal electrons are responsible for the wave excitation. For the self-generation by electrons of momentum $p$, one may replace $\betae^{-1/2}$ by $\pae/p$. This wave generation process is very similar to the low-frequency whistler excitation mechanism driven by an electron heat flux \citep[e.g.,][]{Krafft2010,Roberg-Clark2018,Verscharen2019}. The oblique whistler waves may be destabilized by fast electrons with a large $p/\pae$ resonantly interacting with the wave via the anomalous cyclotron resonance. The consequence of the wave generation is to suppress the upstream-directed electron streaming (i.e., heat flux) through the wave emission in the same direction, which is exactly what is needed for SSDA. The excitation of oblique waves by the shock-reflected electrons (sometimes referred to as electron firehose instability) and the electron scattering associated with the waves observed in PIC simulations may be related to the self-generation mechanism discussed here \citep[e.g.,][]{Matsukiyo2011,Guo2014a,Guo2014b}.

On the other hand, \citet{Amano2010} considered another wave excitation mechanism, in which quasi-parallel high-frequency whistler-mode waves are excited. The instability is driven by an upstream-directed beam of electrons with a loss cone that naturally arises if the beam is produced by magnetic mirror reflection at the shock (or by classical SDA). The electron cyclotron resonance around the loss cone destabilizes the waves propagating downstream. This mechanism may also provide means to induce scattering of upstream-directed electrons. The condition for the instability is roughly given by $\MAHTF \gtrsim (m_i/m_e)^{1/2} \betae^{1/2}$ \citep{Amano2010}. The $\betae$ dependence here appears to overcome the cyclotron damping by thermal electrons. Noting that we have ignored numerical factors of order unity in the threshold Mach numbers, we think both of them give very similar conditions at $\betae \sim 1$. Furthermore, the threshold Mach numbers are consistent with statistical analysis of Earth's bow shock observations (where $\betae \sim 1$ on average), in which harder electron spectra were preferentially found when $\MAHTF \gtrsim (m_i/m_e)^{1/2}$ \citep{Oka2006}. It is thus possible to conjecture that, once the Mach number goes beyond the threshold, quasi-parallel high-frequency whistlers are generated first to trigger SSDA at the lowest energy, and moderately accelerated electrons then excite oblique low-frequency whistlers to realize the continuous particle acceleration over a wide range of energy. We should mention that the self-generation mechanisms originally considered the wave generation in the upstream region to trigger DSA. In reality, however, the wave growth time may be sufficiently short to amplify the wave within the shock transition layer substantially. We thus expect that they should play a crucial role in the acceleration of low-energy electrons through SSDA.

There are certainly possibilities other than the self-generation mechanism discussed above. The high-frequency whistlers may also be generated from adiabatically heated electrons with $T_{\perp}/T_{\parallel} > 1$ that will emit waves in both parallel and anti-parallel directions. However, in-situ observations indicate that quasi-parallel whistlers waves sometimes exhibit uni-directional propagation. This may be due to the quasi-static cross-shock potential that accelerates the bulk of electrons toward the downstream along the local magnetic field line, which makes the electron distribution asymmetric. This will result in the emission of whistlers in the upstream direction \citep{Tokar1984}, which, therefore, cannot scatter the upstream-directed electrons. The low-frequency wave generation by modified two-stream instability (MTSI) favors $\MAHTF \lesssim (m_i/m_e)^{1/2}$ \citep{Matsukiyo2003a,Matsukiyo2006}, which is the opposite of the self-generation model. Although we almost always find intense wave activities around $\omega \sim \omega_{\rm LH}$ in observations, the wave excitation mechanism might be different depending on the shock parameters. Another candidate is lower-hybrid drift instability (LHDI), which may be almost always unstable. Nevertheless, since LHDI is driven by the intense cross-field current at the shock ramp, it may not be a favorable candidate for the low-frequency waves often found in the foot region where the magnetic field gradient is relatively small.

\subsection{Relation to Electron Heating}
\label{heating}
Although our discussion in this paper focuses almost exclusively on the acceleration of supra-thermal electrons, it is closely related to the heating of thermal electrons. It is known that the electron heating at supercritical collisionless shocks is often less efficient compared to ion heating. The reason for this has been considered that electrons behave adiabatically in the shock structure determined by the ion dynamics \citep{Goodrich1984}. In this scenario, the major contribution to electron heating is provided by the cross-shock electrostatic potential and associated microscopic plasma instabilities \citep{Thomsen1987a,Scudder1986a,Schwartz1988b}. More specifically, the field-aligned component of the electric field in HTF accelerates the bulk of electrons toward the downstream. The accelerated electrons may drive electrostatic instabilities, which results in the formation of the so-called flat-top velocity distribution in the downstream \citep{Feldman1982a,Feldman1983a}. The distribution is nearly flat up to the energy corresponding to the shock potential in HTF, which is connected to a power-law spectrum at high energy. Indeed, such a flat distribution, also known as a self-similar distribution, is known to be generated as a result of ion-acoustic turbulence heating \citep{Dum1978a,Dum1978b}. We think that the SSDA theory is responsible for the power-law part of the spectrum beyond the potential energy, which is typically of the order of $\sim 100 \, {\rm eV}$ at Earth's bow shock. This indicates that the electron heating provides the seed population for SSDA. Modeling the electron heating is thus crucial for the overall normalization of the electron spectrum.

Strictly speaking, our model under the diffusion approximation is not able to deal with the low-energy thermal population with the plasma-frame velocity below the HTF shock speed. Such particles can diffuse in the plasma rest frame. However, as they are essentially {\it subsonic}, they cannot propagate back upstream against the fast plasma flow. The diffusion approximation is clearly inadequate for them. Instead, we should adopt the pitch-angle diffusion equation \eqref{eq:kinetic-transport} with the added effect of the parallel electric field associated with the cross-shock potential and turbulent components. This allows us to model the entire electron spectrum with a given source spectrum provided as the boundary condition in the upstream. Although the computational cost for solving the transport equation increases substantially, we do not need to assume any artificial injection into the system in this approach. This is certainly an important subject for future investigation.

It is important to note that recent high-resolution in-situ spacecraft measurements suggested that the heating cannot be understood by the simple classical picture. In other words, reflected-ion-driven or current-driven waves can contribute significantly to the overall heating \citep[e.g.,][]{Wilson2010,Wilson2014,Goodrich2018}. In fact, our injection model implies that even low-energy electrons should somehow behave non-adiabatically. Understanding the roles of large-amplitude waves often observed with in-situ spacecraft is important not only for the heating itself but also for the non-thermal particle production efficiency. Indeed, the electron heating associated with the cross-shock potential alone may not always be sufficient. For the classical SDA to operate in initiating SSDA (and potentially subsequent DSA), the particle must have a sufficiently large pitch-angle defined in HTF. The condition becomes progressively more difficult to satisfy as $\MAHTF$ increases. Since the electron thermal velocity in the solar wind is much larger than the solar wind speed itself, it is not a problem at planetary bow shocks. On the other hand, additional electron heating processes may be needed to trigger SDA at very high Mach number young SNR shocks. It is known that the electrostatic Buneman instability driven by the reflected ion beam may be destabilized in this regime. It was shown that strong energization of electrons via the shock-surfing acceleration (SSA) occurring when they first enter into the shock transition layer may be able to trigger the subsequent SDA \citep{Amano2007,Matsumoto2017}. This suggests that the SSDA may or may not be initiated depending on the heating efficiency through microscopic plasma instabilities.

\subsection{Effect of Non-stationarity}
\label{sec:non-stationarity}

In constructing the simplified theoretical model, we have explicitly made use of HTF, assuming that the shock is stationary. However, it is well known that the collisionless shock exhibits intrinsic non-stationarity. The cyclic self-reformation \citep[e.g.,][]{Leroy1982,Lembege1992a}, shock-surface rippling \citep[e.g.,][]{Winske1988}, and any other types of non-stationarity violates the existence of HTF. It is also noted that a fast-mode MHD shock naturally produces a magnetic field component out of the coplanarity plane within its internal structure \citep{Thomsen1987b}. In such a case, it is impossible to find a frame in which the electric field vanishes throughout the structure even if it is stationary. Considering the highly dynamic and complicated nature of collisionless shocks, one may naturally question the robustness of the electron injection mechanism that assumes the stationarity and the existence of HTF.

We emphasize that the existence of HTF is not an essential ingredient in our model. Our motivation to adopt the specific frame of reference is to simplify the discussion as much as possible, but the application of the particle acceleration mechanism itself should not be limited to a stationary shock. It is known that the particle energization through the classical SDA may be equally understood as the magnetic gradient drift in the direction (anti-parallel direction for electrons) of the motional electric field in NIF \citep{Krauss-Varban1989a}. The use of HTF substantially simplifies the analysis but is not essential. Since the energy gain mechanism of SSDA is the same as SDA, the existence of HTF is not a strict requirement. It does, however, require that the shock is at least locally subluminal. This is because we have assumed that the diffusive particle transport is limited only along the magnetic field line.

It is perhaps important to note that the accelerated electrons will be confined within the shock transition layer for a substantial period of time. It is known that, for the classical SDA, the typical time scale for the electron interaction with the shock structure is comparable to the ion gyroperiod \citep{Krauss-Varban1989b}. In the SSDA mechanism, more efficient confinement via diffusion within the shock transition layer indicates that the interaction time will be much longer. Since the time scale of non-stationarity is typically of the order of the ion gyroperiod, energetic electrons will fully experience the non-stationary shock. It is possible to consider that a non-vanishing electric field associated with the non-stationarity contributes an additional energy gain, but at the same time, the particle confinement efficiency might somewhat deteriorate. While the effect of non-stationarity has not yet been analyzed in detail, there have already been reports on the SSDA signatures with fully kinetic simulations where the shocks are clearly not stationary \citep[][see, Section \ref{sec:simulation} for detail]{Matsumoto2017,Kobzar2021}. This suggests that SSDA can survive even in the non-stationary shock structure.

We conclude that there is always the particle energization mechanism via the magnetic gradient drift, as the magnetic compression is, by definition, an indispensable property of a fast-mode shock. Non-stationary behaviors of the shock may quantitatively affect the particle acceleration efficiency, but not qualitatively. The applicability of the model is rather limited by the magnetic field orientation, which is, however, practically not an issue in non-relativistic shocks.

\subsection{Kinetic Simulations}
\label{sec:simulation}

Fully kinetic PIC simulation has been a powerful tool to investigate the dynamics of collisionless shocks. Recent simulations of non-relativistic oblique shocks have already found signatures of SSDA in the accelerated particle trajectories \citep{Matsumoto2017,Kobzar2021,Ha2021}. More specifically, efficient pitch-angle scattering in momentum space during the energization of particles is qualitatively consistent with the theoretical prediction. On the other hand, the dependence of the spectral index on the shock parameters is not understood. In addition, the spatial dependence of the spectrum has not been carefully analyzed. The theoretical model presented in this paper provides the basis for more quantitative analyses of kinetic simulation results.

Nevertheless, we have to point out the artifacts of numerical simulations. First of all, PIC simulations typically adopt the shock speed one or two orders of magnitude higher than reality. Therefore, even for a modest shock obliquity, the shock may become very close to superluminal. Use of a relativistic HTF shock speed $U_{\rm s}/\cos \thetabn \sim c$ violates the assumption of weak anisotropy made in the present theory. One has to check if the pitch-angle anisotropy for a particular energy range of interest is not substantial before applying the theory. For the nearly isotropic energy range, \eqref{eq:ssda-approx-index} may again be used for calculating the theoretical spectral index.

For more general cases, the effect of pitch-angle anisotropy must be taken into account explicitly by solving \eqref{eq:kinetic-transport}. For instance, we will be able to model energetic particles escaping upstream in a nearly scatter-free manner, which are often seen in simulations of oblique shocks. If the upstream scattering becomes stronger, the escaping particles will be scattered back into the shock to gain further energy in a stochastic manner. Note that such a signature has already been identified in published PIC simulation results and referred to as ``multiple SDA'' \citep[e.g.,][]{Guo2014a,Guo2014b,Ha2021}. In this process, electrons accelerated first at the shock by SDA return back to the shock to gain further energy by another SDA. The particle energy gain thus proceeds without going into the downstream region. As we have seen in this paper, if the pitch-angle anisotropy is sufficiently weak, this process has already been integrated into DSA at an oblique shock. Even though the accelerated particles do not travel deep into the downstream region, the multiple SDA is the signature that is consistent with the particle acceleration by DSA.

Although we have considered only steady-state solutions in this study, time-dependent solutions will also be useful. This allows us to compare the theoretical spectrum with the simulated energy spectrum that may not have reached a steady state.  Theoretically, DSA at an oblique shock will accelerate particles both with SDA and the first-order Fermi process. Typical trajectory analysis, however, tends to pick up for those particles accelerated preferentially by multiple SDA because the first-order Fermi, which requires particles to diffuse into the downstream, is a relatively slow process at a quasi-perpendicular shock. We think that a time-dependent theoretical model will provide a useful guideline for the analysis of particle trajectories obtained by kinetic simulations.

Concerning plasma waves and instabilities in and around collisionless shocks, kinetic simulations have been used extensively over the decades. High-frequency quasi-parallel whistlers, low-frequency oblique whistlers, AIC waves are all identified by PIC simulations in the past \citep[e.g,][]{Umeda2012a,Tran2020,Kobzar2021}. Since they have substantially different frequencies, scale separation by using a large ion-to-electron mass ratio $m_i/m_e$ is important. It should be noted that a PIC simulation of a quasi-perpendicular shock with the realistic mass ratio by \citet{Matsukiyo2015} demonstrated the appearance of all three kinds of waves at the same time. It is interesting to investigate under what conditions those waves of different scales are excited and their roles on particle heating and acceleration.

The use of a large mass ratio is also desirable from the viewpoint of separation of energy scales. Assuming that SSDA starts from the momentum typically gained by the classical SDA $p_1 \sim m_e U_{\rm s} / \cos \thetabn$, we estimate the ratio between the injection threshold $p_{*}$ to the initial momentum $p_1$ as follows
\begin{align}
	\frac{p_{*}}{p_1} \sim
	\left( \frac{k_{*} c}{\wpi} \right)^{-1}
	\left( \frac{m_i}{m_e} \right) \left( \MAHTF \right)^{-1}.
\end{align}
For instance, at Earth's bow shock with $\MA = 8$ and $\thetabn = 85 \degr$, we have $p_{*}/p_1 \approx 20$. This indicates that the electron energy must be increased by a factor of $\approx 400$ for efficient injection. On the other hand, the required energy gain becomes much smaller if a small mass ratio is used instead. For instance, with the same $\MA$ and $\thetabn$, a reduced mass ratio of $m_i/m_e = 100$ often adopted in PIC simulations will make it possible for electrons to interact with ion-scale fluctuations such as the rippling mode even by a single SDA. Therefore, we understand that a small mass ratio artificially makes the electron acceleration more efficient. For a realistic mass ratio, the condition $p_{*}/p_1 \sim 1$ requires $\MAHTF \gtrsim 10^{3}$ or $\MA \gtrsim 10^{2}$ if $\cos \thetabn \sim 0.1$. We have recently shown that the dynamics of such high Mach number shocks will likely be dominated by the Weibel instability rather than the rippling \citep{Nishigai2021}. In this case, whistler waves may no longer be relevant for particle acceleration because electrons can directly interact with the ion-scale turbulence generated by the Weibel instability. The electrons scattered by the large amplitude Weibel-generated turbulence may be accelerated to ultra-relativistic energies within the shock transition layer \citep{Matsumoto2017}. This suggests that the SSDA applies to a wide range of shock parameters.

\subsection{In-situ Observations}
\label{sec:insitu}

While the basis of the theory has already been confirmed by using recent in-situ observations at Earth's bow shock \citep{Amano2020}, the findings in this paper will allow an even more detailed comparison with observations. In particular, the consistency between the scattering rate as a function of energy and the detailed features in the energy spectrum needs to be investigated using the observed profiles of magnetic field and flow velocity. \eqref{eq:general-index-formula} will be particularly useful for comparison between the theoretical and observed spectral slopes for each energy channel of the particle instruments without solving the diffusion-convection equation.

One thing to keep in mind when analyzing the bow shock data is the finite curvature of the shock. Since the accelerated electrons may travel long distances along the magnetic field, the curvature of the shock has a substantial effect on the particle acceleration, especially at high energy. \citet{Krauss-Varban1991} estimated the typical energy beyond which the curvature effect becomes non-negligible for the classical SDA by comparing a ballistic travel distance and the planer scale length. On the other hand, since we consider electrons suffering from pitch-angle scattering, we instead estimate the energy by comparing the diffusion length parallel to the field line $l_{\parallel} = \ldiff/\cos \thetabn$ and the scale size. We assume that the variation of the shock obliquity  $\delta$ as seen by electrons during the particle acceleration is negligible when $l_{\parallel} \lesssim R_{\rm c} \sin \delta \approx R_{\rm c} \delta$, where $R_{\rm c}$ is the curvature radius. This leads to the following condition:
\begin{align}
	E \lesssim {\rm 31 \, keV} \,
	\left( \frac{D_{\mu\mu}/\wce}{0.1} \right)
	\left( \frac{\delta}{\pi/180 {\rm \, rad}} \right)
	\left( \frac{U_{\rm s}}{400 {\rm \, km/s}} \right)
    \nonumber \\ \times
	\left( \frac{B}{10 {\rm \, nT}} \right)
	\left( \frac{R_{\rm c}}{20 R_{\rm E}} \right)
	\left( \frac{\cos \thetabn}{\cos 85 \degr} \right)^{-1},
\end{align}
where $R_{\rm E}$ is the Earth radius. This indicates that, for a typical curvature radius at the bow shock $R_{\rm c} \sim 20 R_{\rm E}$, the scale size of the shock will have a non-negligible effect for electron acceleration beyond a few tens of keV. Note that the diffusive confinement of the particles makes the upper limit much larger than the estimate by \citet{Krauss-Varban1991}. The finite curvature effect may roughly correspond to the presence of the free escaping boundary at a distance of $L_{\rm FEB} \sim R_c \delta \cos \thetabn$ away from the shock. Therefore, under the normal circumstance, Earth's bow shock will not be the place where the electron injection into DSA occurs efficiently. In any case, the high-energy part of the spectrum must be analyzed carefully by taking into account the finite curvature effect.

Investigating the spectrum at high energy around and beyond the transition between SSDA and DSA may be possible at shocks of larger scale sizes such as interplanetary shocks and the bow shocks at Saturn and Jupiter. While interplanetary shocks are favorable in scale size, they are typically much weaker in terms of Mach number. On the other hand, the Saturnian bow shock appears to be a very high Mach number and has a relatively large scale. In one of the shock crossing events observed by Cassini spacecraft, \citet{Masters2013,Masters2017} found that the highest energy range ($\gtrsim 100 \, {\rm keV}$) of the spectrum was harder than the lower energy. We may be able to interpret the hardening of the energy spectrum with the present theory. Although such events are very rare, careful reanalysis might be able to check the consistency with the theory.

Detailed analysis of plasma waves is also desirable. The most important parameter is obviously the intensity of the waves. The condition $\ldiff/L_{\rm s} \lesssim 1$ required for efficient particle acceleration by SSDA indicates that the wave intensity must be larger than a frequency-dependent threshold \citep{Katou2019,Amano2020}. The self-generation hypothesis mentioned above predicts that the {\it absolute} wave intensity will increase as a function of $\MAHTF$. In particular, an abrupt transition across $\MAHTF \sim (m_i/m_e)^{1/2}$ might be realized. On the other hand, even if the wave intensity is kept more or less constant by some other wave excitation mechanisms, the {\it relative} wave intensity normalized to the theoretical threshold will increase as a function of $\MAHTF$ because the threshold is proportional to $(\MAHTF)^{-2}$ if quasi-linear theory applies. Both of the above scenarios will explain the observed correlation between the electron spectral index and $\MAHTF$ found by \citet{Oka2006}. Nevertheless, differentiating the two will provide crucial information for modeling the electron injection.

In addition to the intensity, the wave propagation property is also important. Since quasi-parallel whistlers can scatter low-energy electrons that are moving opposite to the wave propagation direction, their propagation directions must be bi-directional for efficient electron confinement. We have indeed confirmed the bi-directional characteristics of quasi-parallel whistlers in some of the bow shock crossings, but not always. In contrast, the oblique low-frequency whistlers may scatter electrons moving in both directions because of the anomalous cyclotron resonance in addition to the normal cyclotron resonance. If the waves are driven by the upstream-directed electrons, they should always propagate upstream. Since MTSI may generate both upstream and downstream propagating waves, one may distinguish the wave excitation mechanisms using the wave characteristics. While one such attempt has been reported recently \citep{Hull2020}, the dependence on the shock parameters has not been known yet. It is interesting to investigate the possibility that the wave property changes systematically depending on the shock parameters as predicted by the self-generation hypothesis.

\subsection{Laboratory Experiments}
\label{sec:laboratory}

Laboratory experiment using high-power lasers has now become an important tool to study collisionless shocks physics \citep[e.g.,][]{Huntington2015,Schaeffer2017,Schaeffer2019}. Recent development in experimental technologies enables one to perform experiments with various pre-shock plasma magnetizations. In particular, \citet{Fiuza2020} reported non-thermal electron acceleration up to $\sim 500$ keV in Weibel-mediated collisionless shocks reproduced by laser experiments conducted at the National Ignition Facility. A supporting PIC simulation representing the experiment demonstrated a power-law tail in the electron energy spectrum. While they showed that the particle energy gain is consistent with the first-order Fermi mechanism, the spectral index $dN/dE \propto E^{-3}$ or $f(p) \propto p^{-7}$ is clearly steeper than the standard DSA prediction but is more consistent with SSDA. The accelerated particle trajectories indicate that they are confined within the shock transition layer during the energy gain, which implies $\ldiff/L_{\rm s} \lesssim 1$ is satisfied. Note that it may not be appropriate to refer to the particle acceleration mechanism as SSDA for this application because the shock seems to be essentially unmagnetized, in which case SDA arising from the ambient magnetic field compression is absent. Nevertheless, magnetic field compression is not a necessary ingredient for the present particle acceleration mechanism. As long as the scattering is strong enough to confine the particles within the acceleration region, the particle acceleration by the first-order Fermi mechanism in a smooth shock profile may produce a power-law-like spectrum that is steeper than DSA.

Theoretically, we expect that more strongly magnetized conditions may result in more efficient electron acceleration because of the additional energy gain by SDA. On the other hand, if the magnetization is too strong, the Weibel instability will be suppressed, and the shock will behave essentially like the magnetized Earth's bow shock \citep{Nishigai2021}. It would be interesting to investigate the transition between the classical magnetized shock to a more violent Weibel-dominated shock and how this affects the overall electron acceleration efficiency.

\subsection{Remote-sensing Observations}
\label{sec:remotesensing}

The injection model, preferably calibrated with in-situ observations and kinetic simulations, may apply to predict the amount of CR electrons in astrophysical non-thermal emission sources. However, the biggest obstacle for the astrophysical application is the resolution of observations. The ideal assumption that the upstream medium is homogeneous will not be satisfied in practice. It is rather more realistic to assume that observations reflect a superposition of different upstream conditions. In other words, the efficiency of injection will be determined by averaging over the variation of local shock parameters introduced by upstream turbulence interacting with the shock. We note that the particle acceleration mechanism discussed in this paper is dependent only on the local parameters as long as the shock is locally planar. Therefore, if the statistics of the local shock parameters are given, we will be able to predict the absolute flux of CR electrons as a function of $\MAHTF$ using the present injection model.

Observed radio and X-ray synchrotron spectra may be compared directly with the theoretical prediction. While synchrotron photons are emitted from highly relativistic electrons (GeV for radio and TeV for X-ray), we can consistently construct a theoretical spectrum from sub-relativistic to relativistic energies to calculate a theoretical synchrotron spectrum. Another potential probe is non-thermal bremsstrahlung emission in the hard X-ray band from non-relativistic suprathermal electrons. Since the hard X-rays are directly emitted typically from the electrons in the energy range where SSDA plays the dominant role, such observations will provide an important clue for testing the electron injection model. For instance, recent {\it NuSTAR} observation of W49B indicates a power-law-like photon spectrum that was interpreted as non-thermal bremsstrahlung emission \citep{Tanaka2018}. Future X-ray observations of SNRs may provide more detailed information on the electron spectrum in the non-relativistic energy range, which will allow us to perform a direct comparison with the theoretical model.

\section{Summary}
\label{sec:summary}

This paper extends the theory of stochastic shock drift acceleration (SSDA). The mechanism is likely to be responsible for the electron acceleration seen at Earth's bow shock. Furthermore, theoretical parameter dependence indicates that it may explain the injection of electrons into diffusive shock acceleration (DSA) at higher Mach number astrophysical shocks such as young SNR shocks.

SSDA predicts a spectrum steeper than that expected from the standard DSA. Nevertheless, it does not require particles to bounce back and forth across the shock. The particles may be scattered and energized within the shock transition layer where the wave power at high frequency is most intense. Therefore, it is an efficient mechanism for accelerating low-energy electrons. If the diffusion length of accelerated particles becomes longer than the shock thickness, particles may no longer be efficiently trapped inside the shock transition layer. This imposes the upper limit of particle energy obtained through SSDA. If this maximum energy becomes sufficiently high, the particles escaping out from the shock transition layer may be injected into DSA. Although the threshold energy for the electron injection depends on the environment surrounding the shock, the typical energy is on the order of $0.1\text{--}1$ MeV in typical interstellar and interplanetary media. If the threshold energy is reached, the steep spectrum produced by SSDA at low energy may be smoothly connected to the harder spectrum at high energy, which  corresponds to the standard DSA. In general, the injection scenario proposed in this paper favors quasi-perpendicular shocks. More specifically, the \Alfven Mach number in the de Hoffmann-Teller frame (HTF) $\MAHTF = \MA / \cos \thetabn$ is the most significant parameter controlling the efficiency of injection. Higher $\MAHTF$ shocks will be more efficient sites of nonthermal electron acceleration.

Further investigation using observations (both in-situ and remote-sensing) as well as fully kinetic simulations will be needed to make the model more reliable. Rapidly developing laboratory experiments using high-power laser facilities may also be helpful. Ultimately, the theoretical model will be able to predict the absolute flux of CR electrons as a function of macroscopic shock parameters such as the Mach number or magnetic obliquity.

\appendix
\section{Transport Equation}
\label{sec:transport-equation}
We describe the particle transport equation that provides the basis of this paper. We will always assume the presence of HTF (where the motional electric field vanishes) and consider the transport equation in this particular frame of reference (see, Section \ref{sec:non-stationarity} for discussion on the validity). \eqref{eq:general-kinetic-transport} is a fully relativistic pitch-angle diffusion equation both in terms of individual particle speeds and the HTF shock speed. Note that \eqref{eq:general-kinetic-transport} is equivalent to the relativistic transport equation in a general non-inertial frame given by Eq.~(8.5) of \citet{Webb1985,Webb1987}, if it is written in HTF. In this paper, we will restrict ourselves to a non-relativistic shock speed and assume that the pitch-angle anisotropy is weak. The familiar diffusion-convection equation \eqref{eq:diffusion-transport-appendix} obtained in this limit. In the following, we will consistently use notations and reference frames defined in Section \ref{sec:electron-injection}.

\subsection{Relativistic Pitch-angle Diffusion Equation}
\label{sec:pitch-anlge-diffusion}

Let us start with the relativistic Vlasov equation describing the evolution of phase-space density $f(\bm{x}, \bm{p}, t)$:
\begin{eqnarray}
	\frac{\partial f}{\partial t} +
	\frac{\bm{p}}{\gamma m} \cdot \nabla f +
	q \left( \bm{E} + \frac{\bm{p}}{\gamma m c} \times \bm{B} \right) \cdot
	\frac{\partial f}{\partial \bm{p}} = 0,
\end{eqnarray}
for a particle species with the charge $q$, and the rest mass $m$. The momentum is defined by $\bm{p} = \gamma m \bm{v}$ with the particle three-velocity $\bm{v}$ and Lorentz factor $\gamma = 1/\sqrt{1 - v^2/c^2}$. The pitch-angle diffusion equation, also known as the focused transport equation, may be obtained by taking gyrophase average of the Vlasov equation and later introducing a phenomenological pitch-angle diffusion term on the right hand side, assuming that the gyroradius is small compared to the scale size of the inohomogeneity of the electromagnetic field \citep[e.g.,][]{Skilling1975,Isenberg1997}. We will follow the same procedure in HTF. In other words, we assume that the plasma flow $\bm{V}$ is everywhere parallel to the local magnetic field: $\bm{V} = V \bm{b}$, where $\bm{b} = \bm{B}/B$ denotes the magnetic field unit vector. In this case, the convection electric field $\bm{E} = - \bm{V} \times \bm{B} / c$ is always zero. Ignoring also the field-aligned electric field $\bm{E} \cdot \bm{b} = 0$, we obtain
\begin{eqnarray}
	\frac{\partial f}{\partial t} +
	\frac{p \mu}{\gamma m} \bm{b} \cdot \nabla f +
	\frac{1-\mu^2}{2} \frac{p}{\gamma m} \nabla \cdot \bm{b}
	\frac{\partial f}{\partial \mu} = 0,
	\label{eq:guiding-center-vlasov}
\end{eqnarray}
where $p = |\bm{p}|$ and $\mu$ denotes pitch-angle cosine. The absence of a term proportional to $\partial f/\partial p$ clearly indicates that the particle energy is a conserved quantity. Rewriting $\nabla \cdot \bm{b} = - (\bm{b} \cdot \nabla) \ln B$, we see that the transport described by \eqref{eq:guiding-center-vlasov} is nothing more than the adiabatic magnetic mirroring in an inhomogeneous magnetic field.

In \eqref{eq:guiding-center-vlasov}, both the configuration and velocity space coordinates are represented in HTF. However, it is customary and more convenient to use the momentum coordinates defined in the local plasma rest frame, in which momentum space diffusion via wave-particle interaction can be written in the simplest form. By using the Lorentz transform from HTF $(p, \mu)$ to the local plasma rest frame coordinates $(p', \mu')$
\begin{align}
	\gamma c &= \Gamma \left( \gamma' c + p' \mu' V \right),
	\\
	p \mu  &= \Gamma \left( p' \mu' + \gamma' V \right),
\end{align}
where $\Gamma = 1/\sqrt{1 - V^2/c^2}$, \eqref{eq:guiding-center-vlasov} may be rewritten as follows
\begin{align}
	\left( 1 + \frac{v' \mu' V}{c^2} \right) \frac{\partial f}{\partial t}
	& +
	\left( V + v' \mu' \right) \frac{\partial f}{\partial s}
	\nonumber \\
	& -
	\frac{1 - \mu'^2}{2}
	\left[
		(v' + V \mu') \frac{\partial \ln B}{\partial s} +
		2 \Gamma^2 \left( \mu' + \frac{V}{v'} \right)
		\frac{\partial V}{\partial s}
	\right] \frac{\partial f}{\partial \mu'}
	\nonumber \\
	& +
	\left[
		\frac{1-\mu'^2}{2} V \frac{\partial \ln B}{\partial s} -
		\Gamma^2 \left( \mu' + \frac{V}{v'} \right) \mu'
		\frac{\partial V}{\partial s}
	\right] \frac{\partial f}{\partial \ln p'}
	\nonumber \\
	& = \Gamma \frac{\partial}{\partial \mu'}
	\left[ (1 - \mu'^2) D_{\mu\mu} \frac{\partial}{\partial \mu'} f \right].
	\label{eq:general-kinetic-transport}
\end{align}
Note that $\partial/\partial s = \bm{b} \cdot \nabla$ represents the spatial derivative along the local magnetic field line. On the right-hand side, we have introduced a pitch-angle scattering of particles. The scattering conserves energy in the plasma rest frame, and the coefficient $D_{\mu\mu}$ is defined as the rate of scattering measured in that frame. The Lorentz factor $\Gamma$ on the right-hand side indicates the relativistic time dilation effect seen in the laboratory frame.

Henceforth, we will always use the momentum coordinates defined in the local plasma rest frame and drop primes unless otherwise noted. We define $x$ as the coordinate parallel to the normal vector of a plane shock surface, with the positive direction pointing toward the downstream. The angle between the magnetic field and the $x$ axis is denoted by $\theta = \theta (x)$ (thus $\theta(x) = \thetabn$ at $x \rightarrow -\infty$). We may now rewrite Eq.~(\ref{eq:general-kinetic-transport}) in the following form:
\begin{align}
	\left( 1 + \frac{v \mu V}{c^2} \right) \frac{\partial f}{\partial t}
	& +
	\left( V + v \mu \right) \cos \theta \frac{\partial f}{\partial x}
	\nonumber \\
	& -
	\frac{1 - \mu^2}{2}
	\left[
		(v + V \mu) \frac{\partial \ln B}{\partial x} +
		2 \Gamma^2 \left( \mu + \frac{V}{v} \right)
		\frac{\partial V}{\partial x}
	\right] \cos \theta \frac{\partial f}{\partial \mu}
	\nonumber \\
	& +
	\left[
		\frac{1-\mu^2}{2} V \frac{\partial \ln B}{\partial x} -
		\Gamma^2 \left( \mu + \frac{V}{v} \right) \mu
		\frac{\partial V}{\partial x}
	\right] \cos \theta \frac{\partial f}{\partial \ln p}
	\nonumber \\
	& =
	\Gamma \frac{\partial}{\partial \mu}
	\left[ (1 - \mu^2) D_{\mu\mu} \frac{\partial}{\partial \mu} f \right]
	\label{eq:kinetic-transport}
\end{align}
where we have used $\partial/\partial s = \cos \theta \, \partial / \partial x$. This equation may be used to investigate the particle acceleration at plane oblique shocks for arbitrary shock speeds and particle energies, as long as the shock is subluminal. We note that the transport equation used by \citet{Katou2019} may be obtained if both the particle velocity and the shock speed are non-relativistic ($v \ll c$ and $V \ll c$).

\subsection{Diffusion Approximation}
\label{sec:diffusion-approximation}
If the scattering is so strong that the pitch-angle distribution is always kept nearly isotropic, one can further reduce the transport equation to a simpler form. The weak-anisotropy assumption requires that individual particle velocities are much larger than the flow speed $V^2/v^2 \ll 1$, because otherwise, substantial anisotropy will develop in the upstream region. Therefore, a necessary (but not sufficient) condition for the weak-anisotropy approximation is that the shock speed in HTF is non-relativistic $U_{\rm s}/\cos\thetabn \ll c$. While the anisotropy of electrons at suprathermal energies ($\gtrsim 200$ eV) was shown to be weak by in-situ observations at Earth's bow shock \citep{Amano2020}, its validation, in general, requires the knowledge of $D_{\mu\mu}$ even for non-relativistic shock speeds. We here simply adopt this approximation, but the application should be limited to beyond an unspecified threshold energy where the anisotropy becomes weak. The general case with finite pitch-angle anisotropy will be left for future investigation.

Under the assumption of weak anisotropy, we may expand the pitch-angle distribution in terms of the Legendre polynomials:
\begin{align}
	f(x, p, \mu) = \sum_{n=0}^{N} f_n (x, p) P_n (\mu),
\end{align}
where $P_n(\mu)$ is the $n$-th order Legendre polynomial with argument $\mu$. We truncate the expansion to the first order, assuming $|f_0| \gg |f_1|$. This leads to the following diffusion-convection equation for the isotropic part of the distribution fuction $f_0(x,p)$ \citep[see, e.g.,][]{Zank2014b}
\begin{align}
	\frac{\partial f_0}{\partial t} +
	V \cos \theta \, \frac{\partial f_0}{\partial x} +
	\frac{1}{3}
	\left(
		\frac{\partial \ln B}{\partial x} - \frac{\partial \ln V}{\partial x}
	\right)
	V \cos \theta \frac{\partial f_0}{\partial \ln p} =
	\frac{\partial}{\partial x}
	\left(
		\kappa \cos^2 \theta \, \frac{\partial f_0}{\partial x}
	\right).
	\label{eq:diffusion-transport-appendix}
\end{align}
The diffusion coefficient along the ambient magnetic field is given by $\kappa = v^2/6 D_{\mu\mu}$, where we have assumed that $D_{\mu\mu}$ is independent of $\mu$. It is instructive to note
\begin{align}
	\frac{\partial}{\partial x} \left( V \cos \theta \right) =
	- V \cos \theta \left(
		\frac{\partial \ln B}{\partial x} - \frac{\partial \ln V}{\partial x}
	\right),
\end{align}
which indicates
\begin{align}
\frac{1}{p} \frac{d p}{d t} = - \frac{1}{3} \nabla \cdot \bm{V}.
\end{align}
In other words, the particle momentum gain is proportional to the divergence of the plasma flow. We thus find that \eqref{eq:diffusion-transport-appendix} is equivalent to the standard diffusion-convection equation written in HTF.

\citet{Katou2019} showed that the particle energy gain at a nearly perpendicular shock with $\cos \thetabn \ll 1$ is dominated by the gradient-$B$ term.  This may easily be understood by recognizing that the compression of the flow is only in the direction normal to the shock surface, which is only a small fraction of the HTF shock speed. Therefore, we understand that $|\partial \ln B/\partial x| \gg |\partial \ln V/\partial x|$ is satisfied at a quasi-perpendicular shock. Physically, the energy gain represented by $\partial \ln B / \partial x$ is due to SDA, or the gradient-$B$ drift in the direction of the convection electric field. On the other hand, $\partial \ln V / \partial x$ represents the first-order Fermi mechanism by converging scattering centers.

\section{Numerical Method}
\label{sec:appendix-method}
We consider solving the diffusion-convection equation \eqref{eq:diffusion-transport} in the steady-state ($\partial/\partial t = 0$) with a given boundary condition. It is straightforward to solve the equation as an initial-value problem and wait for the time-dependent solution to reach the steady-state after a long time. While it is quite simple and easy to implement in a numerical code, it is not an efficient approach. Furthermore, a large number of grid points should be needed for a large spatial domain (the domain considered in this paper is infinite $-L_{\rm FEB} \leq x < +\infty$). Therefore, we consider an alternative approach that can directly and more efficiently obtain the steady-state solution.

For numerical efficiency, we introduce $\tilde{f} (x, p) = p^{q_0} f(x, p)$ with an arbitrary constant $q_0$. Substituting this into the original equation \eqref{eq:diffusion-transport}, we have the following equation for $\tilde{f} (x, p)$:
\begin{align}
	V_x \frac{\partial \tilde{f}}{\partial x} -
	\frac{1}{3} \frac{\partial V_x}{\partial x}
	\left( \frac{\partial \tilde{f}}{\partial y} - q_0 \tilde{f} \right)
	=
	\frac{\partial}{\partial x}
	\left(
		\kappa_{xx} \frac{\partial \tilde{f}}{\partial x}
	\right),
	\label{eq:appendix-1}
\end{align}
where $y = \ln p$ is the momentum coordinate that will be used in the following instead of $p$. The computational domain in $y$ is thus given by $Y_1 \leq y \leq Y_2$ with $Y_1 = \ln p_1$ and $Y_2 = \ln p_2$. The parameter $q_0$ has been introduced for scaling the numerical solution. For instance, choosing $q_0 = 4$ for a nearly discontinuous shock (in the strong shock limit $r = 4$) will make the solution constnat in $y$. An appropriate choice of $q_0$ thus allows us to obtain an accurate solution with less numerical costs. Note that we consider an injection as the boundary condition at $y = Y_1$. There is no need to specify the boundary condition at $y = Y_2$.

We note that the problem to be solved is essentially a linear boundary-value problem. It should thus be possible to formulate it using a matrix. The question is how to choose the matrix that gives an accurate numerical solution. As mentioned in Section \ref{sec:numerical-solutions}, we divide the domain into three regions and use analytic solutions \eqref{eq:analytic-solution-outside} for the upstream ($-L_{\rm FEB} \leq x \leq X_1$) and downstream ($X_2 \leq x$) regions. This minimizes the degrees of freedom needed to represent the external solutions in the large spatial domains. The standard DSA solution may be obtained by connecting these two solutions directly using appropriate boundary conditions at a discontinuous shock. On the other hand, we have to consider a solution inside the shock transition layer in between the two. Since, by definition, it is not spatially uniform, the solution cannot be represented as a simple analytic form in general.

For the shock transition layer, we expand the solution using general orthogonal basis functions $\Lambda_i (x)$, $\Psi_j (y)$:
\begin{align}
	\tilde{f}(x, p) = \sum_{i=0}^{N_x-1} \sum_{j=0}^{N_y-1} C_{i,j} \Lambda_i (x) \Psi_j (y),
	\label{eq:appendix-2}
\end{align}
where $C_{i,j}$ are the unknown coefficients to be determined. The orders of expansion $N_x$ and $N_y$ are the degrees of freedom used in $x$ and $y$ directions, respectively. Although the basis functions can be chosen arbitrarily as long as they satisfy the orthogonality condition, the choice of these functions is crucial for the accuracy of numerical solutions. We have found that reasonably accurate numerical solutions are obtained with Chebyshev polynomials: $T_n(\xi) = \cos (n \tau)$ with $\xi = \cos \tau$, where $n=0, 1, \ldots$ is the order of polynomials. More specifically, we take $\Lambda_i(x) = T_i(\xi_x)$ and $\Psi_j(y) = T_j(\xi_y)$ where $\xi_x = \left( 2 x - X_2 - X_1 \right) / (X_2 - X_1)$ and $\xi_y = \left( 2 y - Y_2 - Y_1 \right)/(Y_2 - Y_1)$. Note that appropriate scaling of the domains both in $x$ and $y$ are needed here because Chebyshev polynomials are defined in the interval $-1 \leq \xi \leq +1$. Readers more interested in the details of the spectral methods may refer \citet{Boyd2000}.

To determine the unknowns, one has to evaluate the differential equation and the boundary condition using \eqref{eq:appendix-2}. Although the Galarkin method is, in general, known to be slightly more accurate, we use the pseudospectral method instead beause it is much easier to program. For this, we first define a set of collocation points $x_{l} = ((X_2 - X_1) \cos \tau_{x,l} + (X_2 + X_1))/2$ and $y_{m} = ((y_2 - y_1) \cos \tau_{y,m} + (y_2 + y_1))/2$ where
\begin{align}
	& \tau_{x, l} = \frac{l \pi}{N_x - 1}, \quad (l = 0, 1, \ldots, N_x-1)
	\\
	& \tau_{y, m} = \frac{m \pi}{N_y - 1}. \quad (m = 0, 1, \ldots, N_y-1)
\end{align}
Note that $l = 0$ ($m = 0$) and $l = N_x-1$ ($m = N_y-1$) corresponds to $x = X_2$ ($y = Y_2$) and $x = X_1$ ($y = Y_1$), respectively. Substituting \eqref{eq:appendix-2} into \eqref{eq:appendix-1} and evaluating it at the collocation points $(x_l, y_m)$, we obtain
\begin{align}
	\sum_{i=0}^{N_x-1} \sum_{j=0}^{N_y-1}
	& \left[
	\left( V_{x,l} - \kappa'_{xx,l,m} \right)
	\Lambda'_i (x_l) \Psi_j (y_m) -
	\kappa_{xx,l,m}
	\Lambda''_i (x_l) \Psi_j (y_m)
	\right.
	\nonumber \\
	& \left.
	- \frac{1}{3} V'_{x,l}
	\left(
		\Lambda_i (x_l) \Psi'_j (y_m) - q_0 \Lambda_i (x_l) \Psi_j (y_m)
	\right)
	\right] C_{i,j} = 0,
	\label{eq:appendix-matrix}
\end{align}
where $V_{x,l} = V_x (x_l), V'_{x,l} = V'_{x} (x_l), \kappa_{xx,l,m} = \kappa_{xx} (x_l, e^{y_m}), \kappa'_{xx,l,m} = \kappa'_{xx} (x_l, e^{y_m})$ (the prime for $\kappa'_{xx}$ indicates the spatial derivative). The quantities inside the square blacket can easily be evaluated once functional forms for $V_x (x)$ and $\kappa_{xx} (x, p)$ are given. Note that derivatives of $\Lambda_i(x_l)$ and $\Psi_j(y_m)$ should be calculated by evaluating the analytic derivatives at the collocation points. We evaluate \eqref{eq:appendix-matrix} for $l = 1, \ldots, N_x-2$, $m = 0, \ldots, N_y-2$, i.e., except for $x = X_1, X_2$ and $y = Y_1$. The boundary condition at $y = y_1$ given by \eqref{eq:boundary-y1} may be written as
\begin{align}
	\sum_{i=0}^{N_x-1} \sum_{j=0}^{N_y-1}
	\Lambda_i(x_l) \Psi_j(y_1) C_{i,j} =
	f_{\rm inj} \exp \left( - \frac{x_l^2}{2 \sigma^2} \right).
\end{align}
Notice that the number of equations given so far is $(N_x - 2)N_y$, which is less than $N_x N_y$, the number of coefficients $C_{i,j}$. We thus have to consider the boundary conditions at $x = X_1$ and $x = X_2$ to provide additional constraints.

The boundary condition at $x = X_1$ has to connect the solutions defined in the upstream and shock transition layer. We use the continuity of the density $f$ and the flux $\kappa_{xx} \partial f/\partial x$, which are respectively written as
\begin{align}
	& \sum_{i=0}^{N_x-1} \sum_{j=0}^{N_y-1}
	\Lambda_i(X_1) \Psi_j(y_m) C_{i,j} =
	\left[
		\exp
		\left( \dfrac{U_{\rm s} X_1}{\kappa^{\UP}_{xx,m}} \right)
		-
		\exp
		\left(-\dfrac{U_{\rm s} L_{\rm FEB}}{\kappa^{\UP}_{xx,m}} \right)
	\right]
	C^{\UP}_{m},
	\\
	& \sum_{i=0}^{N_x-1} \sum_{j=0}^{N_y-1}
	\Lambda'_i(X_1) \Psi_j(y_m) C_{i,j} =
	\frac{U_{\rm s}}{\kappa_{xx,0,m}}
	\exp \left( \dfrac{U_{\rm s} X_1}{\kappa^{\UP}_{xx,m}} \right)
	C^{\UP}_{m},
\end{align}
where $C^{\UP}_{m} = C^{\UP} (e^{y_m})$ and $\kappa^{\UP}_{xx,m} = \kappa^{\UP} (e^{y_m}) \cos^2 \theta_1$. Note that $\kappa_{xx,0,m}$ may differ from $\kappa^{\UP}_{xx,m}$ in general. Similarly, for the boundary condition at $x = X_2$, we have
\begin{align}
	& \sum_{i=0}^{N_x-1} \sum_{j=0}^{N_y-1}
	\Lambda_i(X_2) \Psi_j(y_m) C_{i,j} =
	C^{\DOWN}_{m},
	\\
	& \sum_{i=0}^{N_x-1} \sum_{j=0}^{N_y-1}
	\Lambda'_i(X_2) \Psi_j(y_m) C_{i,j} =
	0,
\end{align}
where $C^{\DOWN}_m = C^{\DOWN} (e^{y_m})$. The number of equations provided by the boundary conditions is $4 N_y$, while the number of unknowns for the upstream and downstream $C^{\UP}_m$, $C^{\DOWN}_m$ is $2 N_y$. In total, we have the same number of unknowns and equations $(N_x + 2) N_y$. Therefore, the resulting linear matrix equation will have a unique solution that can easily be obtained numerically.

A clear advantage of this method is that it avoids wasting computational resources for the solutions in the upstream and downstream regions, where the analytic solution is available. In particular, for $\ldiff/L_{\rm s} \gg 1$ where the solution is close to the standard DSA, an accurate numerical solution is easily obtained with a small number of degrees of freedom for the internal solution because no complicated internal structure should appear. Since it directly obtains a solution by inverting the matrix, it will take much less computational time than the time-dependent approach for problems in which phenomena of very different time scales coexist. For instance, the Bohm-like diffusion coefficient will make the acceleration time proportional to the momentum. Therefore, a substantial number of time steps will be needed for a problem of a large domain in momentum $p_1 \ll p_2$. The proposed approach is free from such a problem.

\section{Multiscale Plasma Waves}
\label{sec:multiscale-wave}

We present a detailed discussion on plasma waves of three characteristic frequency ranges observed in the vicinity of the shock in the context of electron acceleration by SSDA. As discussed in Section \ref{sec:self-generation}, these multiscale plasma waves play the role in the electron acceleration from slightly above the thermal energy to the injection threshold energy.

The highest frequency range of interest for SSDA is roughly $0.1 \lesssim \omega/\wce \lesssim 0.5$, which corresponds to quasi-parallel whistler-mode waves according to in-situ observations \citep[e.g.,][]{Zhang1999,Hull2012,Oka2017}. These waves can resonantly scatter low-energy electrons slightly above the thermal energy ($\sim$ 0.1-1 keV at the bow shock) \citep{Oka2017,Katou2019,Amano2020}. Therefore, we think that the high-frequency whistlers will play a role in initiating the electron acceleration by SSDA. The high-frequency and coherent nature of the waves indicates that they are probably generated by instability, which may be destabilized by electron cyclotron resonance. In the classical picture, high-frequency whistlers may be generated by temperature anisotropy $T_{\perp}/T_{\parallel} > 1$ or a loss-cone distribution \citep{Tokar1984,Amano2010}. We should mention that recent very high time resolution in-situ measurements by MMS found that the electron velocity distribution function within the shock transition layer exhibits very complicated structures and is highly variable over a short period of time \citep{Chen2018}. Therefore, the actual wave generation mechanism may not necessarily fit into such conventional pictures. Nevertheless, it is important to point out that the stability of the wave is determined by a local slope (in velocity space) of the velocity distribution function around the resonance velocity \citep{Kennel1966c,Gendrin1981}, even though the global distribution does not look like familiar analytic velocity distributions.

The wave intensity is, in general, much higher in the medium frequency range with $\omega \sim \omega_{\rm LH}$. The waves in this frequency range are considered as also on the whistler-mode branch but with propagation angles oblique to the ambient magnetic field \citep{Hull2012,Oka2019,Hull2020}. Since the wave frequency is much lower than the quasi-parallel high-frequency whistlers, they are important candidates for scattering the medium-energy electrons ($\sim$ 1-10 keV at the bow shock). In addition, the oblique propagation angles, as well as larger amplitudes, makes the electron scattering much more efficient \citep{Oka2019}. If SSDA is initiated at lower energy by high-frequency whistlers, moderately accelerated electrons start to interact with the low-frequency oblique whistlers to suffer strong scattering. Because of the larger amplitudes, we expect that the acceleration of medium-energy electrons will be relatively easier. There has been extensive discussion on the origin of the low-frequency oblique whistlers in the shock transition layer. In particular, MTSI and LHDI, which may potentially be coupled with each other, have been believed as the important candidates for the generation mechanism \citep{Hsia1979,Wu1983,Wu1984b,Matsukiyo2003a,Matsukiyo2006}. Both the ion and electron dynamics are involved in these instabilities. In contrast, the self-generation mechanism discussed in Section \ref{sec:self-generation} is driven by sufficiently strong field-aligned electron heat flux, which is based solely on the electron dynamics.

To realize electron injection via SSDA, the electron acceleration must continue roughly up to the energy at which they can resonate with the ion-scale fluctuations $k_{*} c/\wpi \sim 1$. Such high-energy electrons may be scattered by the so-called rippling mode at a quasi-perpendicular shock, which is thought to be generated by temperature anisotropy induced by the shock-reflected ions \citep[e.g.,][]{Winske1988}. The resulting wave mode is essentially a left-hand polarized  low-frequency ($\omega \lesssim \wci$) AIC wave. High-energy electrons near the injection threshold energy may satisfy the cyclotron resonance condition with the AIC waves. Since the ripping mode may have even larger amplitudes, it may help boost the maximum energy of SSDA. We note that if there are ion-scale fluctuations both in the upstream and downstream, the scattering by the rippling mode is not absolutely necessary for injection. Nevertheless, considering that SSDA is a much faster process than DSA, especially at low energies, the additional scattering near the shock will enhance the overall particle acceleration rate.

\acknowledgments
This work was supported by JSPS KAKENHI Grant Nos.~17H02966 and 17H06140. T.~A. is grateful to T.~Katou for the discussion of the transport equation.

\bibliography{reference}{}

\begin{thebibliography}{}
\expandafter\ifx\csname natexlab\endcsname\relax\def\natexlab#1{#1}\fi
\providecommand{\url}[1]{\href{#1}{#1}}
\providecommand{\dodoi}[1]{doi:~\href{http://doi.org/#1}{\nolinkurl{#1}}}
\providecommand{\doeprint}[1]{\href{http://ascl.net/#1}{\nolinkurl{http://ascl.net/#1}}}
\providecommand{\doarXiv}[1]{\href{https://arxiv.org/abs/#1}{\nolinkurl{https://arxiv.org/abs/#1}}}

\bibitem[{Amano \& Hoshino(2007)}]{Amano2007}
Amano, T., \& Hoshino, M. 2007, Astrophys. J., 661, 190, \dodoi{10.1086/513599}

\bibitem[{Amano \& Hoshino(2009)}]{Amano2009a}
---. 2009, Astrophys. J., 690, 244, \dodoi{10.1088/0004-637X/690/1/244}

\bibitem[{Amano \& Hoshino(2010)}]{Amano2010}
---. 2010, Phys. Rev. Lett., 104, 181102,
  \dodoi{10.1103/PhysRevLett.104.181102}

\bibitem[{Amano {et~al.}(2020)Amano, Katou, Kitamura, Oka, Matsumoto, Hoshino,
  Saito, Yokota, Giles, Paterson, Russell, {Le Contel}, Ergun, Lindqvist,
  Turner, Fennell, \& Blake}]{Amano2020}
Amano, T., Katou, T., Kitamura, N., {et~al.} 2020, Phys. Rev. Lett., 124,
  065101, \dodoi{10.1103/PhysRevLett.124.065101}

\bibitem[{Anderson(1981)}]{Anderson1981}
Anderson, K.~A. 1981, J. Geophys. Res., 86, 4445,
  \dodoi{10.1029/JA086iA06p04445}

\bibitem[{Bell(1978)}]{Bell1978a}
Bell, A.~R. 1978, Mon. Not. R. Astron. Soc., 182, 147.
\newblock \url{http://ads.nao.ac.jp/abs/1978MNRAS.182..147B}

\bibitem[{Blandford \& Eichler(1987)}]{Blandford1987}
Blandford, R., \& Eichler, D. 1987, Phys. Rep., 154, 1,
  \dodoi{10.1016/0370-1573(87)90134-7}

\bibitem[{Blandford \& Ostriker(1978)}]{Blandford1978a}
Blandford, R.~D., \& Ostriker, J.~P. 1978, Astrophys. J., 221, L29,
  \dodoi{10.1086/182658}

\bibitem[{Bohdan {et~al.}(2021)Bohdan, Pohl, Niemiec, Morris, Matsumoto, Amano,
  Hoshino, \& Sulaiman}]{Bohdan2021}
Bohdan, A., Pohl, M., Niemiec, J., {et~al.} 2021, Phys. Rev. Lett., 126,
  095101, \dodoi{10.1103/PhysRevLett.126.095101}

\bibitem[{Boyd(2000)}]{Boyd2000}
Boyd, J.~P. 2000, {Chebyshev and Fourier Spectral Methods}, 688.
\newblock
  \url{http://books.google.com/books?hl=en{\&}lr={\&}id=lEWnQWyzLQYC{\&}oi=fnd{\&}pg=PR10{\&}dq=Chebyshev+and+Fourier+Spectral+Methods{\&}ots=WRdKyCo9nt{\&}sig=IO1U70ic8UM3-CDhwYegnCuh7fE}

\bibitem[{Chen {et~al.}(2018)Chen, Wang, Wilson, Schwartz, Bessho, Moore,
  Gershman, Giles, Malaspina, Wilder, Ergun, Hesse, Lai, Russell, Strangeway,
  Torbert, F.-Vinas, Burch, Lee, Pollock, Dorelli, Paterson, Ahmadi, Goodrich,
  Lavraud, {Le Contel}, Khotyaintsev, Lindqvist, Boardsen, Wei, Le, \&
  Avanov}]{Chen2018}
Chen, L.-J., Wang, S., Wilson, L.~B., {et~al.} 2018, Phys. Rev. Lett., 120,
  225101, \dodoi{10.1103/PhysRevLett.120.225101}

\bibitem[{Dresing {et~al.}(2016)Dresing, Theesen, Klassen, \&
  Heber}]{Dresing2016}
Dresing, N., Theesen, S., Klassen, A., \& Heber, B. 2016, Astron. Astrophys.,
  588, A17, \dodoi{10.1051/0004-6361/201527853}

\bibitem[{Drury(1983)}]{Drury1983}
Drury, L.~O. 1983, Reports Prog. Phys., 46, 973,
  \dodoi{10.1088/0034-4885/46/8/002}

\bibitem[{Drury(1991)}]{Drury1991}
---. 1991, Mon. Not. R. Astron. Soc., 251, 340.
\newblock \url{http://adsabs.harvard.edu/abs/1991MNRAS.251..340D}

\bibitem[{Drury {et~al.}(1982)Drury, Axford, \& Summers}]{Drury1982}
Drury, L.~O., Axford, W.~I., \& Summers, D. 1982, Mon. Not. R. Astron. Soc.,
  198, 833, \dodoi{10.1093/mnras/198.3.833}

\bibitem[{Dum(1978{\natexlab{a}})}]{Dum1978a}
Dum, C.~T. 1978{\natexlab{a}}, Phys. Fluids, 21, 945, \dodoi{10.1063/1.862338}

\bibitem[{Dum(1978{\natexlab{b}})}]{Dum1978b}
---. 1978{\natexlab{b}}, Phys. Fluids, 21, 956, \dodoi{10.1063/1.862339}

\bibitem[{Feldman {et~al.}(1982)Feldman, Bame, Gary, Gosling, McComas, Thomsen,
  Paschmann, Sckopke, Hoppe, \& Russell}]{Feldman1982a}
Feldman, W., Bame, S., Gary, S., {et~al.} 1982, Phys. Rev. Lett., 49, 199,
  \dodoi{10.1103/PhysRevLett.49.199}

\bibitem[{Feldman {et~al.}(1983)Feldman, Anderson, Bame, Gary, Gosling,
  McComas, Thomsen, Paschmann, \& Hoppe}]{Feldman1983a}
Feldman, W.~C., Anderson, R.~C., Bame, S.~J., {et~al.} 1983, J. Geophys. Res.,
  88, 96, \dodoi{10.1029/JA088iA01p00096}

\bibitem[{Fiuza {et~al.}(2020)Fiuza, Swadling, Grassi, Rinderknecht, Higginson,
  Ryutov, Bruulsema, Drake, Funk, Glenzer, Gregori, Li, Pollock, Remington,
  Ross, Rozmus, Sakawa, Spitkovsky, Wilks, \& Park}]{Fiuza2020}
Fiuza, F., Swadling, G.~F., Grassi, A., {et~al.} 2020, Nat. Phys., 16, 916,
  \dodoi{10.1038/s41567-020-0919-4}

\bibitem[{Gendrin(1981)}]{Gendrin1981}
Gendrin, R. 1981, Rev. Geophys., 19, 171, \dodoi{10.1029/RG019i001p00171}

\bibitem[{Goodrich \& Scudder(1984)}]{Goodrich1984}
Goodrich, C.~C., \& Scudder, J.~D. 1984, J. Geophys. Res., 89, 6654,
  \dodoi{10.1029/JA089iA08p06654}

\bibitem[{Goodrich {et~al.}(2018)Goodrich, Ergun, Schwartz, Wilson, Newman,
  Wilder, Holmes, Johlander, Burch, Torbert, Khotyaintsev, Lindqvist,
  Strangeway, Russell, Gershman, Giles, \& Andersson}]{Goodrich2018}
Goodrich, K.~A., Ergun, R., Schwartz, S.~J., {et~al.} 2018, J. Geophys. Res.,
  123, 9430, \dodoi{10.1029/2018JA025830}

\bibitem[{Gosling {et~al.}(1989)Gosling, Thomsen, Bame, \&
  Russell}]{Gosling1989}
Gosling, J.~T., Thomsen, M.~F., Bame, S.~J., \& Russell, C.~T. 1989, J.
  Geophys. Res., 94, 10011, \dodoi{10.1029/JA094iA08p10011}

\bibitem[{Guo {et~al.}(2014{\natexlab{a}})Guo, Sironi, \& Narayan}]{Guo2014a}
Guo, X., Sironi, L., \& Narayan, R. 2014{\natexlab{a}}, Astrophys. J., 794,
  153, \dodoi{10.1088/0004-637X/794/2/153}

\bibitem[{Guo {et~al.}(2014{\natexlab{b}})Guo, Sironi, \& Narayan}]{Guo2014b}
---. 2014{\natexlab{b}}, Astrophys. J., 797, 47,
  \dodoi{10.1088/0004-637X/797/1/47}

\bibitem[{Ha {et~al.}(2021)Ha, Kim, Ryu, \& Kang}]{Ha2021}
Ha, J.-H., Kim, S., Ryu, D., \& Kang, H. 2021, Astrophys. J., 915, 18,
  \dodoi{10.3847/1538-4357/abfb68}

\bibitem[{Hoshino \& Shimada(2002)}]{Hoshino2002}
Hoshino, M., \& Shimada, N. 2002, Astrophys. J., 572, 880,
  \dodoi{10.1086/340454}

\bibitem[{Hsia {et~al.}(1979)Hsia, Chiu, Hsia, Chou, \& Wu}]{Hsia1979}
Hsia, J.~B., Chiu, S.~M., Hsia, M.~F., Chou, R.~L., \& Wu, C.~S. 1979, Phys.
  Fluids, 22, 1737, \dodoi{10.1063/1.862810}

\bibitem[{Hull {et~al.}(2020)Hull, Muschietti, {Le Contel}, Dorelli, \&
  Lindqvist}]{Hull2020}
Hull, A.~J., Muschietti, L., {Le Contel}, O., Dorelli, J.~C., \& Lindqvist, P.
  2020, J. Geophys. Res., 125, \dodoi{10.1029/2019JA027290}

\bibitem[{Hull {et~al.}(2012)Hull, Muschietti, Oka, Larson, Mozer, Chaston,
  Bonnell, \& Hospodarsky}]{Hull2012}
Hull, A.~J., Muschietti, L., Oka, M., {et~al.} 2012, J. Geophys. Res., 117,
  A12104, \dodoi{10.1029/2012JA017870}

\bibitem[{Huntington {et~al.}(2015)Huntington, Fiuza, Ross, Zylstra, Drake,
  Froula, Gregori, Kugland, Kuranz, Levy, Li, Meinecke, Morita, Petrasso,
  Plechaty, Remington, Ryutov, Sakawa, Spitkovsky, Takabe, \&
  Park}]{Huntington2015}
Huntington, C.~M., Fiuza, F., Ross, J.~S., {et~al.} 2015, Nat. Phys., 11, 173,
  \dodoi{10.1038/nphys3178}

\bibitem[{Isenberg(1997)}]{Isenberg1997}
Isenberg, P.~A. 1997, J. Geophys. Res., 102, 4719, \dodoi{10.1029/96JA03671}

\bibitem[{Johlander {et~al.}(2016)Johlander, Schwartz, Vaivads, Khotyaintsev,
  Gingell, Peng, Markidis, Lindqvist, Ergun, Marklund, Plaschke, Magnes,
  Strangeway, Russell, Wei, Torbert, Paterson, Gershman, Dorelli, Avanov,
  Lavraud, Saito, Giles, Pollock, \& Burch}]{Johlander2016}
Johlander, A., Schwartz, S.~J., Vaivads, A., {et~al.} 2016, Phys. Rev. Lett.,
  117, 165101, \dodoi{10.1103/PhysRevLett.117.165101}

\bibitem[{Katou \& Amano(2019)}]{Katou2019}
Katou, T., \& Amano, T. 2019, Astrophys. J., 874, 119,
  \dodoi{10.3847/1538-4357/ab0d8a}

\bibitem[{Kennel \& Petschek(1966)}]{Kennel1966c}
Kennel, C.~F., \& Petschek, H.~E. 1966, J. Geophys. Res., 71, 1,
  \dodoi{10.1029/jz071i001p00001}

\bibitem[{Kobzar {et~al.}(2021)Kobzar, Niemiec, Amano, Hoshino, Matsukiyo,
  Matsumoto, \& Pohl}]{Kobzar2021}
Kobzar, O., Niemiec, J., Amano, T., {et~al.} 2021, Astrophys. J., 919, 97,
  \dodoi{10.3847/1538-4357/ac1107}

\bibitem[{Krafft \& Volokitin(2010)}]{Krafft2010}
Krafft, C., \& Volokitin, A. 2010, Phys. Plasmas, 17, 102303,
  \dodoi{10.1063/1.3479829}

\bibitem[{Krauss-Varban \& Burgess(1991)}]{Krauss-Varban1991}
Krauss-Varban, D., \& Burgess, D. 1991, J. Geophys. Res., 96, 143,
  \dodoi{10.1029/90JA01728}

\bibitem[{Krauss-Varban {et~al.}(1989)Krauss-Varban, Burgess, \&
  Wu}]{Krauss-Varban1989b}
Krauss-Varban, D., Burgess, D., \& Wu, C.~S. 1989, J. Geophys. Res., 94, 15089,
  \dodoi{10.1029/JA094iA11p15089}

\bibitem[{Krauss-Varban \& Wu(1989)}]{Krauss-Varban1989a}
Krauss-Varban, D., \& Wu, C.~S. 1989, J. Geophys. Res., 94, 15367,
  \dodoi{10.1029/ja094ia11p15367}

\bibitem[{Lario(2003)}]{Lario2003}
Lario, D. 2003, {ACE Observations of Energetic Particles Associated with
  Transient Interplanetary Shocks}, Vol. 679 (AIP), 640--643,
  \dodoi{10.1063/1.1618676}

\bibitem[{Lembege \& Savoini(1992)}]{Lembege1992a}
Lembege, B., \& Savoini, P. 1992, Phys. Fluids B, 4, 3533,
  \dodoi{10.1063/1.860361}

\bibitem[{Leroy \& Mangeney(1984)}]{Leroy1984a}
Leroy, M.~M., \& Mangeney, A. 1984, Ann. Geophys., 2, 449.
\newblock \url{http://adsabs.harvard.edu/abs/1984AnGeo...2..449L}

\bibitem[{Leroy {et~al.}(1982)Leroy, Winske, Goodrich, Wu, \&
  Papadopoulos}]{Leroy1982}
Leroy, M.~M., Winske, D., Goodrich, C.~C., Wu, C.~S., \& Papadopoulos, K. 1982,
  J. Geophys. Res., 87, 5081, \dodoi{10.1029/JA087iA07p05081}

\bibitem[{Levinson(1992)}]{Levinson1992a}
Levinson, A. 1992, Astrophys. J., 401, 73, \dodoi{10.1086/172039}

\bibitem[{Levinson(1996)}]{Levinson1996}
---. 1996, Mon. Not. R. Astron. Soc., 278, 1018.
\newblock \url{http://adsabs.harvard.edu/abs/1996MNRAS.278.1018L}

\bibitem[{Livesey {et~al.}(1984)Livesey, Russell, \& Kennel}]{Livesey1984}
Livesey, W.~A., Russell, C.~T., \& Kennel, C.~F. 1984, J. Geophys. Res., 89,
  6824, \dodoi{10.1029/JA089iA08p06824}

\bibitem[{Masters {et~al.}(2013)Masters, Stawarz, \& Fujimoto}]{Masters2013}
Masters, A., Stawarz, L., \& Fujimoto, M. 2013, Nat. Phys., 8,
  \dodoi{10.1038/NPHYS2541}

\bibitem[{Masters {et~al.}(2017)Masters, Sulaiman, Stawarz, Reville, Sergis,
  Fujimoto, Burgess, Coates, \& Dougherty}]{Masters2017}
Masters, A., Sulaiman, A.~H., Stawarz, {\L}., {et~al.} 2017, Astrophys. J.,
  843, 147, \dodoi{10.3847/1538-4357/aa76ea}

\bibitem[{Matsukiyo \& Matsumoto(2015)}]{Matsukiyo2015}
Matsukiyo, S., \& Matsumoto, Y. 2015, J. Phys. Conf. Ser., 642, 012017,
  \dodoi{10.1088/1742-6596/642/1/012017}

\bibitem[{Matsukiyo {et~al.}(2011)Matsukiyo, Ohira, Yamazaki, \&
  Umeda}]{Matsukiyo2011}
Matsukiyo, S., Ohira, Y., Yamazaki, R., \& Umeda, T. 2011, Astrophys. J., 742,
  47, \dodoi{10.1088/0004-637X/742/1/47}

\bibitem[{Matsukiyo \& Scholer(2003)}]{Matsukiyo2003a}
Matsukiyo, S., \& Scholer, M. 2003, J. Geophys. Res., 108, 1459,
  \dodoi{10.1029/2003JA010080}

\bibitem[{Matsukiyo \& Scholer(2006)}]{Matsukiyo2006}
---. 2006, J. Geophys. Res., 111, A06104, \dodoi{10.1029/2005JA011409}

\bibitem[{Matsumoto {et~al.}(2017)Matsumoto, Amano, Kato, \&
  Hoshino}]{Matsumoto2017}
Matsumoto, Y., Amano, T., Kato, T.~N., \& Hoshino, M. 2017, Phys. Rev. Lett.,
  119, 105101, \dodoi{10.1103/PhysRevLett.119.105101}

\bibitem[{Nishigai \& Amano(2021)}]{Nishigai2021}
Nishigai, T., \& Amano, T. 2021, Phys. Plasmas, 28, 072903,
  \dodoi{10.1063/5.0051269}

\bibitem[{Oka {et~al.}(2006)Oka, Terasawa, Seki, Fujimoto, Kasaba, Kojima,
  Shinohara, Matsui, Matsumoto, Saito, \& Mukai}]{Oka2006}
Oka, M., Terasawa, T., Seki, Y., {et~al.} 2006, Geophys. Res. Lett., 33, 5,
  \dodoi{10.1029/2006GL028156}

\bibitem[{Oka {et~al.}(2017)Oka, III, Phan, Hull, Amano, Hoshino, Argall,
  Contel, Agapitov, Gershman, Khotyaintsev, Burch, Torbert, Pollock, Dorelli,
  Giles, Moore, Saito, Avanov, Paterson, Ergun, Strangeway, Russell, \&
  Lindqvist}]{Oka2017}
Oka, M., III, L. B.~W., Phan, T.~D., {et~al.} 2017, Astrophys. J., 842, L11,
  \dodoi{10.3847/2041-8213/aa7759}

\bibitem[{Oka {et~al.}(2019)Oka, Otsuka, Matsukiyo, Wilson, Argall, Amano,
  Phan, Hoshino, Contel, Gershman, Burch, Torbert, Dorelli, Giles, Ergun,
  Russell, \& Lindqvist}]{Oka2019}
Oka, M., Otsuka, F., Matsukiyo, S., {et~al.} 2019, Astrophys. J., 886, 53,
  \dodoi{10.3847/1538-4357/ab4a81}

\bibitem[{Reynolds(2008)}]{Reynolds2008}
Reynolds, S.~P. 2008, Annu. Rev. Astron. Astrophys., 46, 89,
  \dodoi{10.1146/annurev.astro.46.060407.145237}

\bibitem[{Roberg-Clark {et~al.}(2018)Roberg-Clark, Drake, Reynolds, \&
  Swisdak}]{Roberg-Clark2018}
Roberg-Clark, G.~T., Drake, J.~F., Reynolds, C.~S., \& Swisdak, M. 2018, Phys.
  Rev. Lett., 120, 035101, \dodoi{10.1103/PhysRevLett.120.035101}

\bibitem[{Schaeffer {et~al.}(2019)Schaeffer, Fox, Follett, Fiksel, Li,
  Matteucci, Bhattacharjee, \& Germaschewski}]{Schaeffer2019}
Schaeffer, D.~B., Fox, W., Follett, R.~K., {et~al.} 2019, Phys. Rev. Lett.,
  122, 245001, \dodoi{10.1103/PhysRevLett.122.245001}

\bibitem[{Schaeffer {et~al.}(2017)Schaeffer, Fox, Haberberger, Fiksel,
  Bhattacharjee, Barnak, Hu, \& Germaschewski}]{Schaeffer2017}
Schaeffer, D.~B., Fox, W., Haberberger, D., {et~al.} 2017, Phys. Rev. Lett.,
  119, 025001, \dodoi{10.1103/PhysRevLett.119.025001}

\bibitem[{Scholer {et~al.}(2003)Scholer, Shinohara, \&
  Matsukiyo}]{Scholer2003a}
Scholer, M., Shinohara, I., \& Matsukiyo, S. 2003, J. Geophys. Res., 108,
  \dodoi{10.1029/2002JA009515}

\bibitem[{Schwartz {et~al.}(1988)Schwartz, Thomsen, Bame, \&
  Stansberry}]{Schwartz1988b}
Schwartz, S.~J., Thomsen, M.~F., Bame, S.~J., \& Stansberry, J. 1988, J.
  Geophys. Res., 93, 12923, \dodoi{10.1029/ja093ia11p12923}

\bibitem[{Scudder {et~al.}(1986)Scudder, Mangeney, Lacombe, Harvey, Aggson,
  Anderson, Gosling, Paschmann, \& Russell}]{Scudder1986a}
Scudder, J.~D., Mangeney, A., Lacombe, C., {et~al.} 1986, J. Geophys. Res., 91,
  11019, \dodoi{10.1029/JA091iA10p11019}

\bibitem[{Shimada \& Hoshino(2000)}]{Shimada2000}
Shimada, N., \& Hoshino, M. 2000, Astrophys. J., 543, L67,
  \dodoi{10.1086/318161}

\bibitem[{Skilling(1975)}]{Skilling1975}
Skilling, J. 1975, Mon. Not. R. Astron. Soc., 172, 557.
\newblock \url{http://adsabs.harvard.edu/full/1975MNRAS.172..557S}

\bibitem[{Tanaka {et~al.}(2018)Tanaka, Yamaguchi, Wik, Aharonian, Bamba,
  Castro, Foster, Petre, Rho, Smith, Uchida, Uchiyama, \&
  Williams}]{Tanaka2018}
Tanaka, T., Yamaguchi, H., Wik, D.~R., {et~al.} 2018, Astrophys. J., 866, L26,
  \dodoi{10.3847/2041-8213/aae709}

\bibitem[{Thomsen {et~al.}(1987{\natexlab{a}})Thomsen, Gosling, Bame, Quest,
  Winske, Livesey, \& Russell}]{Thomsen1987b}
Thomsen, M.~F., Gosling, J.~T., Bame, S.~J., {et~al.} 1987{\natexlab{a}}, J.
  Geophys. Res., 92, 2305, \dodoi{10.1029/JA092iA03p02305}

\bibitem[{Thomsen {et~al.}(1987{\natexlab{b}})Thomsen, Mellott, Stansberry,
  Bame, Gosling, \& Russell}]{Thomsen1987a}
Thomsen, M.~F., Mellott, M.~M., Stansberry, J.~A., {et~al.} 1987{\natexlab{b}},
  J. Geophys. Res., 92, 10119, \dodoi{10.1029/JA092iA09p10119}

\bibitem[{Tokar {et~al.}(1984)Tokar, Gurnett, \& Feldman}]{Tokar1984}
Tokar, R.~L., Gurnett, D.~A., \& Feldman, W.~C. 1984, J. Geophys. Res., 89,
  105, \dodoi{10.1029/JA089iA01p00105}

\bibitem[{Tran \& Sironi(2020)}]{Tran2020}
Tran, A., \& Sironi, L. 2020, Astrophys. J., 900, L36,
  \dodoi{10.3847/2041-8213/abb19c}

\bibitem[{Umeda {et~al.}(2012{\natexlab{a}})Umeda, Kidani, Matsukiyo, \&
  Yamazaki}]{Umeda2012b}
Umeda, T., Kidani, Y., Matsukiyo, S., \& Yamazaki, R. 2012{\natexlab{a}}, Phys.
  Plasmas, 19, 042109, \dodoi{10.1063/1.3703319}

\bibitem[{Umeda {et~al.}(2012{\natexlab{b}})Umeda, Kidani, Matsukiyo, \&
  Yamazaki}]{Umeda2012a}
---. 2012{\natexlab{b}}, J. Geophys. Res., 117, A03206,
  \dodoi{10.1029/2011JA017182}

\bibitem[{Verscharen {et~al.}(2019)Verscharen, Chandran, Jeong, Salem, Pulupa,
  \& Bale}]{Verscharen2019}
Verscharen, D., Chandran, B. D.~G., Jeong, S.-Y., {et~al.} 2019, Astrophys. J.,
  886, 136, \dodoi{10.3847/1538-4357/ab4c30}

\bibitem[{Webb(1985)}]{Webb1985}
Webb, G.~M. 1985, Astrophys. J., 296, 319, \dodoi{10.1086/163451}

\bibitem[{Webb(1987)}]{Webb1987}
---. 1987, Astrophys. J., 321, 606, \dodoi{10.1086/165656}

\bibitem[{Wilson {et~al.}(2010)Wilson, Cattell, Kellogg, Goetz, Kersten,
  Kasper, Szabo, \& Wilber}]{Wilson2010}
Wilson, L.~B., Cattell, C.~A., Kellogg, P.~J., {et~al.} 2010, J. Geophys. Res.,
  115, n/a, \dodoi{10.1029/2010JA015332}

\bibitem[{Wilson {et~al.}(2014)Wilson, Sibeck, Breneman, Contel, Cully, Turner,
  Angelopoulos, \& Malaspina}]{Wilson2014}
Wilson, L.~B., Sibeck, D.~G., Breneman, A.~W., {et~al.} 2014, J. Geophys. Res.,
  119, 6475, \dodoi{10.1002/2014JA019930}

\bibitem[{Wilson {et~al.}(2016)Wilson, Sibeck, Turner, Osmane, Caprioli, \&
  Angelopoulos}]{Wilson2016}
Wilson, L.~B., Sibeck, D.~G., Turner, D.~L., {et~al.} 2016, Phys. Rev. Lett.,
  117, 215101, \dodoi{10.1103/PhysRevLett.117.215101}

\bibitem[{Winske \& Quest(1988)}]{Winske1988}
Winske, D., \& Quest, K.~B. 1988, J. Geophys. Res., 93, 9681,
  \dodoi{10.1029/JA093iA09p09681}

\bibitem[{Wu(1983)}]{Wu1983}
Wu, C.~S. 1983, Phys. Fluids, 26, 1259, \dodoi{10.1063/1.864285}

\bibitem[{Wu(1984)}]{Wu1984b}
---. 1984, J. Geophys. Res., 89, 8857, \dodoi{10.1029/JA089iA10p08857}

\bibitem[{Zank(2014)}]{Zank2014b}
Zank, G.~P. 2014, Lecture Notes in Physics, Vol. 877, {Transport Processes in
  Space Physics and Astrophysics} (New York, NY: Springer New York),
  \dodoi{10.1007/978-1-4614-8480-6}

\bibitem[{Zhang(1999)}]{Zhang1999}
Zhang, Y. 1999, J. Geophys. Res., 104, 449, \dodoi{10.1029/1998JA900049-A}

\end{thebibliography}
\bibliographystyle{aasjournal}



\end{document}